\documentclass[11pt]{article}
\usepackage{amsmath,latexsym,amssymb,epsfig,psfrag}
\newcommand{\be}{\begin{equation}}
\newcommand{\ee}{\end{equation}}

\newcommand{\I}{\mathrm i} 
\newcommand{\ba}{\begin{array}} 
\newcommand{\ea}{\end{array}}
\newcommand{\bal}{\begin{align}}
\newcommand{\eal}{\end{align}}
\newcommand{\bqa}{\begin{eqnarray}}
\newcommand{\eqa}{\end{eqnarray}}
\newcommand{\nms}{\normalsize}

\newcommand{\abs}[1]{\left|#1\right|}

\newcommand{\tam}[1]{{\text {\Large $#1$}}}

\newcommand{\lb}{\left(}
\newcommand{\rb}{\right)}
\newcommand{\lsb}{\left[}
\newcommand{\rsb}{\right]}
\newcommand{\lcb}{\left\{}
\newcommand{\rcb}{\right\}}

\newcommand{\hypF}{\hspace{2mm}F_{\hspace{-3.8mm}1\hspace{1.9mm}1}}
\renewcommand{\lim}[2]{\begin{tabular}{c} \vspace{-3mm}lim\\ $_{#1\to #2}$\end{tabular}}
\def\pd{\partial}
\def\spd{\slash\hspace{-2.1mm}\partial}
\def\di{{\rm d}}
\def\a{\alpha}
\def\b{\beta}

\def\g{\gamma}

\def\G{\Gamma}
\def\d{\delta}

\def\k{{\kappa}}
\def\m{\mu}
\def\n{\nu}

\def\t{\tau}
\def\th{\theta}

\def\th{\theta}
\def\z{\zeta}
\def\fz{\Theta}
\def\l{\lambda}

\def\sig{\sigma}
\def\Sig{\Sigma}
\def\e{\epsilon}

\def\id{{\bf 1}}

\def\H{\mathfrak H}


\begin{document}
\begin{titlepage}  

\vskip 2cm
%\begin{flushright}
%{\bf UAB-FT-XXX}
%\end{flushright}
\begin{center}  
\vspace{0.5cm} \Large {\sc FERMION MASS HIERARCHY FROM THE SOFT WALL}
 %\Large {\sc A possible origin of neutrino masses.}                            
\vspace*{1.5cm}
  
\normalsize  
  
{\bf 
Antonio Delgado~\footnote[1]{antonio.delgado@nd.edu}} and {\bf David Diego~\footnote[2]{david.diego@nd.edu}}
%G.v.~Gersdorff~\footnote{gero@pha.jhu.edu} 
%and
%M.~Quir\'os~\footnote{quiros@ifae.es} 
%}   

\smallskip   
\medskip   
{\it Department of Physics, 225 Nieuwland Science Hall, U. of Notre Dame,}\\ 
{\it Notre Dame, IN 46556-5670, USA.}

\vskip0.6in \end{center}  
   
\centerline{\large\bf Abstract}  

\vspace{.5cm}

\noindent

We develop a 5d model for ElectroWeak physics based on a non compact warped extra dimension of finite length, known as the soft wall scenario, where all the dynamical degrees of freedom propagate in the 5d bulk. 
We solve the equations of motion and find the allowed spectra, showing that the mass of the lightest fermionic mode behaves as a power law of the effective 4d Yukawa coupling constant, with the exponent being the corresponding fermionic 5d bulk mass. 
 Precisely this non universal behavior allows us to reproduce the hierarchy between the Standard Model (SM) fermion masses (from neutrinos to the top quark) with non-hierarchycal fermionic bulk masses. 
 %{\it HERE IT GOES THE ABSTRACT}
   
\vspace*{2mm}   
%\smallskip\newline  
  
\end{titlepage}  
%
%
%
%
%
%%%%%%%%%%%%%%%%%%%%%%%%%%%%%%%%%%%%%%%%%%%%%%%%%%%%%%%%%%%%%%%%%%%%%%%%%%%%%%%%%%%%%%
%%%%%%%%%%%%%%%%%%%%%%%%%%%%%%%%%%%%%%%%%%%%%%%%%%%%%%%%%%%%%%%%%%%%%%%%%%%%%%%%%%%%%%
\section{\Large Introduction}
%%%%%%%%%%%%%%%%%%%%%%%%%%%%%%%%%%%%%%%%%%%%%%%%%%%%%%%%%%%%%%%%%%%%%%%%%%%%%%%%%%%%%
%%%%%%%%%%%%%
  The existence of warped extra dimensions has been a source of prolific investigations for the last 10 years because  it was shown that it could provide with an alternative  
  explanation of the hierarchy between the Planck and the ElectroWeak scales~\cite{Randall:1999vf,antoniadis}. 
  The pioneer work by Randall and Sundrum~\cite{Randall:1999vf} considers a slice of a five dimensional anti de Sitter space spanning 
  between two 3-branes,  the UV one accounts for the gravity (Planckian) scale and the IR brane is chosen to have a scale of the order of the
   ElectroWeak breaking by the exponential suppression of the AdS metric.  Another aspect of this set-up
   that has attracted a lot of attention is its connection to the AdS/CFT duality \cite{adscft} which states that this 5d theory is equivalent to a 4d CFT coupled to gravity and the interpretation of the IR scale is that it corresponds to the breaking of that conformal symmetry and the appearance of a mass gap and resonances \cite{rscft}. 
   
     More recently, an alternative scenario with a (non compact) warped extra dimension has been proposed~\cite{Karch:2006pv,Falkowski:2008fz,Batell:2008me}, the so-called soft wall
  as opposed to the hard wall termination of the Randall-Sundrum model. 
  The motivation for the new scenario, as pointed out in reference~\cite{Karch:2006pv}, comes from the fact that a suitable soft wall configuration may lead to linear Kaluza-Klein (KK) excitations ($m_n\sim n$), unlike the quadratic 
  excitations found in the hard wall models, and hence it may provide a description of the linear spectrum of mesons in QCD~\cite{Csaki:2006ji}. This is what has been called the AdS/QCD approach.

  In the soft wall case the (effective) metric is not AdS, instead the warped factor decays faster enough to make the extra dimension of finite length and in this sense a virtual
   IR boundary is located at infinity but with a finite distance from the UV one and the departure from the AdS behavior is associated to a smooth VEV acquired
   by some dilaton field, however, since no real IR boundary exists all the matter/radiation content propagate in the 5d bulk, thus the 
   suppression due to the metric is universal (and is absorbed through the wave function normalization). The possible suppression on the lowest modes comes from the binding of the 
   Schr\"odinger-like potentials governing the  profiles of the wave functions~\cite{Falkowski:2008fz,Batell:2008me}.
   
   In this work we try to reproduce the hierarchy of the fermion masses  of the Standard Model embedding it  on a soft wall background. It is worth remarking, nevertheless, that
   our main concern here is to reproduce the correct order of magnitude for the masses, a more realistic study taking into account all the families and all the ElectroWeak constrains would need to incorporate the CKM matrices, 
  which we leave for future investigations. In any case, the incorporation of the mixing angles would not change the main results we are presenting. 
  Finally, since the topic is relatively new and relatively absent in the literature we think a detailed derivation, although overlapping previous publications, is worthwhile.
  
    The paper is structured as follows: In section~\ref{themodel} we present the model together with the notations and conventions. Section~\ref{eom} is devoted to solve the equations 
  of motion for the fermion, Higgs and gauge sectors. Section~\ref{phen} applies the previous results to the hierarchy of the fermion masses and section~\ref{conclusions} 
  contains the conclusions. Finally, at the end of the paper, we have included appendices developing some technical aspects of particular calculations.     
%%%%%%%%%%%%%%%%%%%%%%%%%%%%%%%%%%%%%%%%%%%%%%%%%%%%%%%%%%%%%%%%%%%%%%%%%%%%%%%%%%%%%%
%%%%%%%%%%%%%%%%%%%%%%%%%%%%%%%%%%%%%%%%%%%%%%%%%%%%%%%%%%%%%%%%%%%%%%%%%%%%%%%%%%%%%%
\section{\Large The Model}\label{themodel}
%%%%%%%%%%%%%%%%%%%%%%%%%%%%%%%%%%%%%%%%%%%%%%%%%%%%%%%%%%%%%%%%%%%%%%%%%%%%%%%%%%%%%
%%%%%%%%%%%%%
  Before presenting the model we will set up the notation used: Capital letters from the beginning of the alphabet ($A,\,B, \cdots$) represent (local) Lorentz 5d indices 
 while those from the middle and end of the alphabet ($M,\,N,\,R, \cdots$) stand for
 general coordinate 5d indices. Finally, greek letters take into account 4d Lorentz indices.
 
  The 5d Lorentz representation is spanned 
 by the gamma matrices $\g^A = \lb\g^\m,-\I\g^5\rb$, with $\g^\m$ the usual 4d gamma matrices and $$\g^5 = \lb\ba{cc}\id&0\\0&-\id\ea\rb\,.$$  
 
  Our model is defined in a five dimensional manifold with (UV) boundary, $\Sig = \mathcal \mathbb R^4\times\mathcal I$, with $\mathcal I = \left.\lsb z_0, \infty\right.\rb$ and $z_0 > 0$, 
 assuming  also $z_0\ll \lb M_W\rb^{-1}$ being $M_W$ the ElectroWeak scale. The background metric is AdS, that is
  \be
  g = f^2(z)\,\eta_5\,,\label{metric}
  \ee 
 where $f(z) = k/z$, $k$ being the radius of curvature and as a matter of fact we will take $z_0 \sim k$.
 In addition $\eta_5$ stands for the 5d Minkowski metric with mostly negative signature, 
 namely $\eta_{5\,AB} = {\rm diagonal}\lb+1,-1,-1,-1,-1\rb$ and $z$ represents the fifth coordinate. 
 The propagating degrees of freedom consist of an $SU(2)_L$ doublet of 5d Dirac fermions: $\Psi^i_L$ and two singlets $\Psi^\mathfrak a_R$,
  an $SU(2)_L$ doublet of complex scalars, $H^i$,
 and a gauge field, $A_M$, lying in the adjoint representation~\footnote{To avoid important quantum corrections to the $\rho$ parameter the gauge group should be $SU(2)_L\times SU(2)_R\times U(1)_Y$~\cite{Falkowski:2008fz,Batell:2008me}
although for the sake of simplicity we take the usual gauge group. For the calculations of the spectra this technicality does not affect.} of $SU(2)_L\times U(1)_Y$,
where the lightest modes of the above fields are to be identified with the SM fermions, Higgs and ElectroWeak gauge filelds, respectively. For the 5d fields, the bulk-brane 
 action is\footnote{The presence of the (5,5) component of the inverse vielbein, $e^5_5$, in the brane action is not a problem because the (effective) background metric itself breaks the isometry group to 4d Lorentz-Poincar\'e.} 
 \begin{align}
 S &= \int_\Sig\sqrt{g}\,\tam{e}^{-\phi} \lsb \frac{\I}{2}\bar\Psi^i_L\g^M\mathcal D^L_M\Psi^i_L+\frac{\I}{2}\bar\Psi^{\mathfrak a}_R\g^M\mathcal D^R_M\Psi^{\mathfrak a}_R+  M\,\bar\Psi^i_L\Psi^i_L + M\,\bar\Psi^{\mathfrak a}_R\Psi^{\mathfrak a}_R\right.\nonumber\\
    %&\hspace{1cm} \nonumber\\
        %
        % &\hspace{2.6cm}
         &\hspace{1cm} +\lb\l_1\,\e_{ij}\bar\Psi^i_L\Psi^1_R\,H^{*\,j}+\l_2\,H^{i}\bar\Psi^i_L\Psi^2_R + {\rm h.c.}\rb\nonumber\\
         %
         %&\hspace{2.6cm}\left. +
         &\hspace{1cm}\left.+ g^{MN}\,\lb D_M H\rb^\dagger D_N H - m_h^2 H^\dagger H+\frac{1}{4 g_5^2} \,g^{MR} g^{N S}\,{\rm tr}\lcb F_{MN} F_{R S}\rcb\rsb\nonumber\\
         &- \int_{\pd\Sig} \sqrt g \,\tam{e}^{-\phi}\lsb\l_0 k^2 \lb \abs{H}^2- v_0^2\rb^2 - e^5_5\frac{1}{2}\lb\bar\Psi^i_L \Psi^i_L-\bar\Psi^{\mathfrak a}_R\Psi^{\mathfrak a}_R\rb\rsb\,,\label{action}
 \end{align}
 where $F_{MN} = \pd_{\lsb M\right.} A_{\left.N\rsb} + \lsb A_M, A_N\rsb$ and $A_M = g_5\,A_M^s T_s$ $T_s$ being the group generators verifying $\lsb T_s, T_t\rsb = \I f_{st}^r T_r$ 
 and $g_5$ the (5d) gauge coupling constant, $\l^{1,\,2}$ are the 5d Yukawa coupling constants, $M$ is a (constant) bulk mass and 
 $m_h^2 = \frac{a \lb a-4\rb}{k^2} - 2 a \frac{\m^2}{k^2}\, z^2$ with $a, \l_0$ being dimensionless constants whereas $v_0$ has dimension of $\lb{\rm energy}\rb^{3/2}$. 
 $\e_{ij}$ is the Levi-Civita tensor and $\phi$ is the dilaton field. In addition
 \begin{align}
 &D_M = \pd_M + \I A_M\,,\\
   &\mathcal D^L_M = D_M + E_M^C\G^A_{CB}\eta_{AD} \Sig^{BD} \,,\\
   &\mathcal D^R_M = \pd_M + E_M^C\G^A_{CB}\eta_{AD} \Sig^{BD} \,,
 \end{align}
 are gauge and Lorentz covariant derivatives,
 where $E_M^A$ is the vielbein field, verifying $g_{MN}=E_M^A E_N^B \,\eta_{5\,AB}$. The gamma matrices depend on the spacetime coordinates according to
 $\g^M = e^M_A \g^A$ with $e^M_A$ being the inverse of the vielbein field verifying $g_{MN}\,e^M_A e^N_B =\eta_{5\,AB}$, i.e. they are the (local coordinate) components of orthonormal vector 
 fields ($e_A \equiv e_A^M \pd_M$).
 Furthermore, $\Sig^{AB} = \frac{1}{8} \lsb\g^A,\g^B\rsb$ are the  
 Lorentz group generators in the representation 
 $${\rm diagonal}\lb SL(2,\mathbb C)\times SL(2,\mathbb C)^{\dagger\,-1}\rb\,,$$ 
 and the symbols $\G^A_{BC}$ verify
 $\nabla_{e_B} e_C = \G^A_{BC}\,e_A$ with $\nabla$ the Levi-Civita connection compatible with $g$. The relation with the Christoffel symbols, $ \G^R_{MN}$,
 is then $\G^A_{BC} = e_B^M E^A_N \pd_M e_C^N + e_B^M e_C^N E_R^A\,\G^R_{MN}$. In the case of the metric (\ref{metric}) the Lorentz covariant derivative is
 \begin{align}
   &\mathcal D^R_\m = \pd_\m + \frac{\I}{2}\,\g^5\g_\m\frac{f'}{f} \,,\\
   &\mathcal D^R_5 = \pd_5 \,.
 \end{align}
 
 For the dilaton field we will assume a $z$-dependent VEV. The particular bulk-brane gravitational dynamic leading to that profile for the dilaton is beyond 
 the scope of the present work, although in Ref.~\cite{Batell:2008me} an underlying gravity model leading to a soft wall configuration can be found. Here we will take
 $\langle\phi\rangle = \m^2 z^2$ with $\m$ some mass scale. As for the Higgs bulk mass is concerned, its $z$ dependence can be understood as a consequence of the coupling to the gravity sector. We will also not address the question of the different scales in the problem but supposed that their stability is granted by an underlaying UV theory.
 
 Finally, we have taken the same ($z$-independent) bulk mass for Left- and Right-handed fermions for the sake of simplicity at solving the equations of motion. Different masses, as it can be 
 seen in the next section, substantially increase the difficulty as it is not possible to decouple easily the first order fermionic equations of motion.   
  %%%%%%%%%%%%%%%%%%%%%%%%%%%%%%%%%%%%%%%%%%%%%%%%%%%%%%%%%%%%%%%%%%%%%%%%%%%%%%%%%%%%%%
%%%%%%%%%%%%%%%%%%%%%%%%%%%%%%%%%%%%%%%%%%%%%%%%%%%%%%%%%%%%%%%%%%%%%%%%%%%%%%%%%%%%%%
\section{\Large Solving the equations of motion}\label{eom}
%%%%%%%%%%%%%%%%%%%%%%%%%%%%%%%%%%%%%%%%%%%%%%%%%%%%%%%%%%%%%%%%%%%%%%%%%%%%%%%%%%%%%
%%%%%%%%%%%%%
We will start by finding the VEV of the Higgs following the derivation found in Ref.~\cite{Batell:2008me}. 
 %%%%%%%%%%%%%%%%%%%%%%%%%%%%%%%%%%%%%%%%%%%%%%%%%%%%%%%%%%%%%%%%%%%%%%%%%%%%%%%%%%%%%%
%%%%%%%%%%%%%%%%%%%%%%%%%%%%%%%%%%%%%%%%%%%%%%%%%%%%%%%%%%%%%%%%%%%%%%%%%%%%%%%%%%%%%%
\subsection{\Large Higgs VEV}
\label{higgsvev}
%%%%%%%%%%%%%%%%%%%%%%%%%%%%%%%%%%%%%%%%%%%%%%%%%%%%%%%%%%%%%%%%%%%%%%%%%%%%%%%%%%%%%
%%%%%%%%%%%%%
Consider the purely scalar sector\footnote{We consider the Higgs doublet in the unitary gauge, that is $H = \lb\ba{c} 0\\ \H\ea\rb$ with $\H\in\mathbb R$.} out of the action (\ref{action})
and assume for the moment $H(x,z) \equiv \H(z)$, then, the minimal action principle yields the equations of motion
\be
\H''+\lb3\frac{f'}{f} - \phi'\rb\,\H' - f^2 m_h^2\,\H =0\,,\label{bulk}
\ee
together with the boundary conditions
\be
\left.\H' - 2\l_0 k^2 \lb \H^2- v_0^2\rb \H\right|_{z_0}=0\,.\label{bound}
\ee
Redefining the scalar as $\H = z^a \mathcal H$ we find the equation
\be
\mathcal H''+\lb\frac{2 a -3}{z} - 2 \m^2 z\rb\,\mathcal H' =0\,,
\ee
and thus the general solution is
\be
   \H = z^a \lb c + \tilde c\, \int^z_{z_0}\di\z\, \z^{3-2 a}\, \tam{e}^{\m^2 \z^2}\rb\,,
\ee
although the normalizable solution is obtained with $\tilde c=0$ and hence the boundary conditions set the constraint
\be
  c\lsb a - 2\l_0 k^2 \lb c^2 z_0^{2\,a}- v_0^2\rb z_0\rsb =0\,,
\ee
whose solution is
\be
\lcb\ba{l} c =0\\ \\ c^2 = \lb\frac{1}{z_0}\rb^{2a}\lb\frac{a}{ 2 z_0 \l_0 k^2} + v_0^2\rb\ea\right.
\ee
Now if we compute the energy density per unit 4d volume~\cite{Coradeschi:2007gb} for the solutions to the system (\ref{bulk})-(\ref{bound}) we find 
 \begin{align}
 \mathcal E_{\H} =& \int_\Sig\sqrt{g}\,\tam{e}^{-\phi}  \lb -g^{55}\pd_5 \H\pd_5\H + m_h^2 \H^2\rb + \int_{\pd\Sig} \sqrt g \,\tam{e}^{-\phi} \l_0 k^2 \lb \H^2- v_0^2\rb^2\nonumber\\
 %\end{align}
 %
%\begin{align}
= & \int_\Sig \sqrt g \,\tam{e}^{-\phi} \H \lsb g^{55}\nabla_5\pd_5\H -g^{55}\pd_5\phi\,\pd_5 \H + m^2_h\, \H \rsb\nonumber\\
  &+ \int_{\pd\Sig} \sqrt g \,\tam{e}^{-\phi}\lsb \H\,\H' +\l_0 k^2 \lb\H^2- v_0^2\rb^2\rsb\nonumber\\
  =& \l_0 k^2 \int_{\pd\Sig} \sqrt g \,\tam{e}^{-\phi}\lb v_0^4- \H^4 \rb\,.
\end{align}
 The energy density per unit 4d volume is then decreasing around the trivial solution, $\H =0$, for $\l_0 >0$ and therefore 
in this case the non trivial solution is the absolute minimum. 

Now that the Higgs profile has been set, we then proceed by solving the equations of motion for the fermion sector according to three polynomial cases for the Higgs VEV.
 %%%%%%%%%%%%%%%%%%%%%%%%%%%%%%%%%%%%%%%%%%%%%%%%%%%%%%%%%%%%%%%%%%%%%%%%%%%%%%%%%%%%%%
%%%%%%%%%%%%%%%%%%%%%%%%%%%%%%%%%%%%%%%%%%%%%%%%%%%%%%%%%%%%%%%%%%%%%%%%%%%%%%%%%%%%%%
\subsection{\Large Fermionic sector}
\label{fersec}
\begin{figure}[tb]
\begin{center}
\includegraphics[width=8cm]{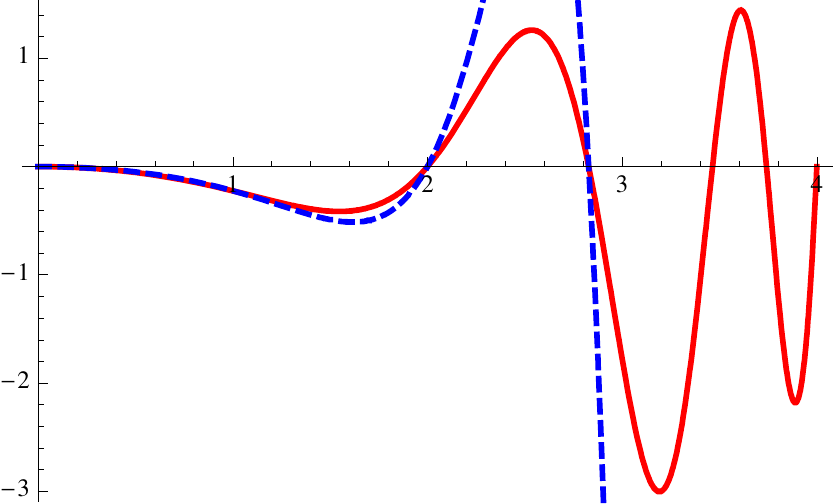}
\caption{\it Normalized fermionic mass quantization condition as a function of $\d$ for $\z_0 = 10^{-3}$ and $\a = 3$ (solid line) and $\a = 5$ (dashed line).}
\label{quant}
\end{center}
\end{figure}

%%%%%%%%%%%%%%%%%%%%%%%%%%%%%%%%%%%%%%%%%%%%%%%%%%%%%%%%%%%%%%%%%%%%%%%%%%%%%%%%%%%%%
%%%%%%%%%%%%%
%%%%%%%%%%%%%%%%%%%%%%%%
%%%%%%%%%%%%%%%%%%%%%%%%
%%%%%%%%%%%%%%%%%%%%%%%%
Applying the variational principle to the fermionic sector\footnote{Considering the Lorentz covariant derivative only.} of the action (\ref{action}) one finds a bulk piece whose vanishing induces the following equations of motion  
 \begin{align}
   &\I e^M_A\g^A\mathcal D^R_M \Psi^i_L - \frac{1}{2} e^5_A\g^A \phi' \Psi^i_L + M\,\Psi^i_L + \l_1 \e_{ij}\H^{j}\,\Psi^1_R+ \l_2 \H^{i}\,\Psi^2_R  = 0\,,\label{eqmL}\\
   &\I e^M_A\g^A\mathcal D^R_M \Psi^1_R - \frac{1}{2} e^5_A\g^A \phi' \Psi^1_R + M\,\Psi^1_R + \l_1 \e_{ij}\Psi^i_L\H^{j} = 0\,,\\\label{eqmR}
   &\I e^M_A\g^A\mathcal D^R_M \Psi^2_R - \frac{1}{2} e^5_A\g^A \phi' \Psi^2_R + M\,\Psi^2_R + \l_2 \H^{i}\,\Psi^i_L = 0
 \end{align}
 while the boundary variation yields the constraint 
 \be
    e^5_5 \lb\id-\sig_3\otimes\g^5\rb\lb\ba{c}\Psi^i_L\\ \Psi^{\mathfrak a}_R\ea\rb_{z_0} =0\,.
 \ee
 where $\sig_3$ is the $z$-Pauli matrix acting on $\lb L, R\rb$ space. If we refer $\Psi^i_L$ and $\Psi^{\mathfrak a}_R$ as simply $\Psi_L$ and $\Psi_R$, respectively, 
 the equations of motion and the boundary conditions adopt the generic form 
  \begin{align}
   &\I e^M_A\g^A\mathcal D^R_M \Psi_L - \frac{1}{2} e^5_A\g^A \phi' \Psi_L + M\,\Psi_L + \l\, \H(z)\,\Psi_R \,,\\
   &\I e^M_A\g^A\mathcal D^R_M \Psi_R - \frac{1}{2} e^5_A\g^A \phi' \Psi_R + M\,\Psi_R + \l\, \H(z)\,\Psi_L = 0\,,\\
   &e^5_5 \lb\id-\sig_3\otimes\g^5\rb\lb\ba{c}\Psi_L\\ \Psi_R\ea\rb_{z_0} =0\,.
\end{align}
 and upon the redefinitions\footnote{With different $L$ and $R$ fermionic bulk masses one can not decouple the system with a global rotation.}
 \be\Psi_{L,R} = \frac{1}{\sqrt 2} f^{-2} \tam{e}^{\frac{1}{2}\phi}\lb\psi_+ \mp\psi_-\rb\,,\label{red}\ee 
 we find 
  \be
    \I\spd\,\psi_\pm+ \g^5\pd_5\psi_\pm + \lb M \pm m_D\rb f\,\psi_\pm=0\,,\label{1storder}
 \ee
 with $m_D = \l\,\H(z)$, while the boundary conditions take the form 
 \be
   \lb\id-\sig_1\otimes\g^5\rb\lb\ba{c}\psi_+\\\psi_-\ea\rb = 0\,.\label{0bc}
 \ee
 Considering now an orbifold-like decomposition\footnote{Notice that the normalization of the kinetic term in these variables is simply $\int^{\infty}_{z_0} \abs{h}^2 $.}, that is
 \be
     \psi_\pm = \lb\ba{c} h_\pm(z) \xi_\pm(x^\m)\\ g_\pm(z) \bar\chi_\pm(x^\m)\ea\rb\,,
 \ee
 where $\xi_\pm, \chi_\pm$ are Weyl spinors verifying $\I\bar\sig^\m\pd_\m\xi_\pm = m\,\bar\chi_\pm$, $m$ being the physical eigenmass, the 
 first order coupled differential equations (\ref{1storder}) can be written as 
 \begin{align}
    & h'_\pm + \lb M \pm m_D\rb f\, h_\pm = - m \,g_\pm\,,\label{firsth}\\
    & g'_\pm - \lb M \pm m_D\rb f\, g_\pm =  m \,h_\pm\,,\label{firstg}
 \end{align}
  and differentiating once more they transform into two decoupled 2nd order differential equations
 \begin{align}
   & h''_\pm - \lsb f^2 \lb M \pm m_D\rb^2 - f' \lb M\pm m_D\rb \mp f\,m'_D - m^2\rsb\,h_\pm = 0\,,\label{secondh}\\
   & g''_\pm - \lsb f^2 \lb M \pm m_D\rb^2 + f' \lb M\pm m_D\rb \pm f\,m'_D - m^2\rsb\,g_\pm = 0\,.\label{secondg}
 \end{align}
 In addition the boundary conditions (\ref{0bc}), in terms of the new variables, translate into\footnote{Eq.~(\ref{0bc}) implies $\chi_+ = \k\chi_-$ with $\k$ a constant (and an analogous relation
 for $\xi_\pm$) however by redefining the Weyl spinors one can absorb the constants.}
 \be
   \lb\id - \sig_1\rb\lb\ba{c}h_+\\ h_-\ea\rb = \lb\id + \sig_1\rb\lb\ba{c}g_+\\ g_-\ea\rb =0\,.\label{bc}
 \ee
 In the following we explicitly solve these equations for three polynomial behaviors of the Higgs VEV.
\begin{itemize}
 \item Constant profile ($a =0$): $\H(z) = v_0$
\end{itemize}     
%%%%%%%%%%%%%%%%%%%%%%%%
%%%%%%%%%%%%%%%%%%%%%%%%
%%%%%%%%%%%%%%%%%%%%%%%%
In this case the equations take the form 
\begin{align}
   & \frac{\di}{\di\,\th}\,h_\pm +  \frac{\a_\pm}{\th}\, h_\pm = - \,g_\pm\,,\label{hpm}\\
    & \frac{\di}{\di\,\th}\,g_\pm -  \frac{\a_\pm}{\th}\, g_\pm =  \,h_\pm\,,\label{gpm}\\
    \nonumber\\
    &\frac{\di^2}{\di\,\th^2}\,h_\pm -\frac{\a_\pm\lb\a_\pm +1\rb}{\th^2} h_\pm +  h_\pm = 0\,,\\
    &\frac{\di^2}{\di\,\th^2}\,g_\pm -\frac{\a_\pm\lb\a_\pm -1\rb}{\th^2} g_\pm +  g_\pm = 0\,,
\end{align}
with {\nms $\a_\pm = M k \pm \l\,v_0 k$} and $\th = m z$. Redefining the fuctions as $h_\pm = \sqrt{\th}\,\mathfrak h_\pm $ and $g_\pm = \sqrt{\th}\,\mathfrak g_\pm $ we find 
the Bessel equations
\begin{align}
  & \th\frac{\di}{\di\th}\lb\th\frac{\di}{\di\th}\,\mathfrak h_\pm\rb + \lsb \th^2 -\lb\a_\pm + \frac{1}{2}\rb^2\rsb\,\mathfrak h_\pm=0\,,\\
  & \th\frac{\di}{\di\th}\lb\th\frac{\di}{\di\th}\,\mathfrak g_\pm\rb + \lsb \th^2 -\lb\a_\pm - \frac{1}{2}\rb^2\rsb\,\mathfrak g_\pm=0\,,
\end{align}
 and, accordingly, the solutions are
 \begin{align}
    &h_\pm = \sqrt{\th}\lsb H_\pm^J\, J\lb\a_\pm+\frac{1}{2}\,,\th\rb + H_\pm^Y\, Y\lb\a_\pm+\frac{1}{2}\,,\th\rb \rsb\,,\\
    &g_\pm = \sqrt{\th}\lsb G_\pm^J \,J\lb\a_\pm-\frac{1}{2}\,,\th\rb + G_\pm^Y \,Y\lb\a_\pm-\frac{1}{2}\,,\th\rb \rsb\,,
 \end{align}
 for $\a_\pm\pm \frac{1}{2}$ non-integer. One can easily check that
 \begin{align}
 & \lb\th\frac{\di}{\di\th} +\a_\pm\rb\lsb\sqrt{\th}\,J\lb\frac{1}{2}+\a_\pm,\th\rb\rsb =\th\sqrt{\th}\,J\lb-\frac{1}{2}+\a_\pm,\th\rb \,,\\
 & \lb\th\frac{\di}{\di\th} +\a_\pm\rb\lsb\sqrt{\th}\,Y\lb\frac{1}{2}+\a_\pm,\th\rb\rsb =\th\sqrt{\th}\,Y\lb-\frac{1}{2}+\a_\pm,\th\rb \,,
 \end{align}
 and hence the first order constraints (\ref{hpm})-(\ref{gpm}) impose
 \begin{align}
 &G^J_\pm  =- H^J_\pm   \,,\\
 &G^Y_\pm  = -H^Y_\pm   \,.
 \end{align}
 Since we have enough degrees of freedom to solve the boundary conditions (\ref{bc}), no further restriction on the mass eigenvalue is found. Thus, the spectrum is continuum and there is no
 mass gap. %
 %%%%%%%%%%%%%%%%%%%%%%%%
%%%%%%%%%%%%%%%%%%%%%%%%
%%%%%%%%%%%%%%%%%%%%%%%%
 \begin{itemize}
 \item Linear profile ($a=1$): $\H(z) = \frac{\tilde\m}{\sqrt k}\, \frac{z}{k}$ with $\tilde\m = \frac{1}{z_0}\sqrt{\frac{k}{ 2 z_0 \l_0} + v_0^2 k^3}$.
 \end{itemize}
 %%%%%%%%%%%%%%%%%%%%%%%%
%%%%%%%%%%%%%%%%%%%%%%%%
%%%%%%%%%%%%%%%%%%%%%%%%
   The equations of motion are now 
   \begin{align}
      &\frac{\di}{\di\fz} h_{\pm} + \lb\frac{M k}{\fz}\pm \frac{\l}{2 \sqrt k}\frac{\tilde\m}{\g}\rb h_{\pm} = -\frac{m}{2 \g} g_{\pm}\,,\label{linear1sth}\\
       &\frac{\di}{\di\fz} g_{\pm} - \lb\frac{M k}{\fz}\pm \frac{\l}{2 \sqrt k}\frac{\tilde\m}{\g}\rb g_{\pm} = \frac{m}{2 \g} h_{\pm}\,,\label{linear1stg}\\
       \nonumber\\
       &\frac{\di^2}{\di\fz^2} h_{\pm} - \lsb\frac{M k \lb Mk + 1\rb}{\fz^2} \pm \frac{M \l \tilde\m \sqrt k}{\g\,\fz} +\frac{1}{4}\rsb h_{\pm } = 0\,,\label{linearhpm}\\
       &\frac{\di^2}{\di\fz^2} g_{\pm} - \lsb\frac{M k \lb Mk - 1\rb}{\fz^2} \pm \frac{M \l \tilde\m \sqrt k}{\g\,\fz} +\frac{1}{4}\rsb g_{\pm } = 0\,, \label{lineargpm}
   \end{align}
   with $\Theta = 2 \g\,z$ and $\g = \sqrt{\frac{\l^2}{k}\tilde\m^2 - m^2}$. Then with the redefinitions 
   \be
     h_{\pm} = \tam{e}^{-\fz/2}\, \fz^{1+ Mk}\, \mathfrak h_{\pm}\,,\qquad g_{\pm} = \tam{e}^{-\fz/2}\, \fz^{1-M k}\, \mathfrak g_{\pm}\,,
   \ee
    Eqs. (\ref{linearhpm})-(\ref{lineargpm}) turn into the following confluent hypergeometric equations (alsol known as Kummer equations)
   \begin{align}
     &\fz\,\frac{\di^2}{\di\fz^2}\,\mathfrak h_{\pm} +\lb2+2\a - \fz\rb \frac{\di}{\di\fz}\,\mathfrak h_{\pm} - \lb 1+\a\pm\b\rb \,\mathfrak h_\pm = 0\,,\\
      &\fz\,\frac{\di^2}{\di\fz^2} \,\mathfrak g_{\pm} +\lb2-2\a - \fz\rb \frac{\di}{\di\fz}\, \mathfrak g_{\pm} - \lb1-\a\pm\b\rb \,\mathfrak g_\pm = 0\,,
   \end{align}
  where $\a = M k$ and $\b = \frac{M \l \tilde\m \sqrt k}{\g}$. For $2\pm 2\a\notin\mathbb Z$ the general solution is (see appendix~\ref{B})
  \begin{align}
  &\mathfrak h_{\pm} = H_{\pm} \hypF\lb 1+\a\pm\b,2+2\a ,\fz\rb + H'_\pm\, U\lb 1+\a\pm\b,2+2\a ,\fz\rb\,,\\
  &\mathfrak g_{\pm} = G_{\pm} \hypF\lb 1-\a\pm\b,2-2\a ,\fz\rb + G'_\pm\, U\lb 1-\a\pm\b,2-2\a ,\fz\rb\,.
  \end{align}
  Furthermore, one easily show that
  \begin{align}
    &\lsb\frac{\di}{\di\fz}+ \lb\frac{\a}{\fz}\pm \frac{\b}{2\a}\rb \rsb\lcb\fz^{1+\a}\,\tam{e}^{-\fz/2}\hypF\lb 1+\a\pm\b,2+2\a,\fz\rb\rcb=\nonumber\\
    & \lb1+ 2 \a\rb \fz^{\a} \,\tam{e}^{-\fz/2}\hypF\lb-\a\pm\b,-2\a,\fz\rb\,,\\
    \nonumber\\
    &\lsb\frac{\di}{\di\fz}+ \lb\frac{\a}{\fz}\pm \frac{\b}{2 \a}\rb \rsb\lcb\fz^{1+\a}\,\tam{e}^{-\fz/2} U\lb 1+\a\pm\b,2+2\a,\fz\rb\rcb=\nonumber\\
    & \frac{-\a\pm\b}{2\a} \,\fz^{\a}\, \tam{e}^{-\fz/2} \,U\lb-\a\pm\b,-2\a,\fz\rb\,,
     \end{align}
     and so for $M \to -M$ and $h\to g$, 
  thus the general solution to the coupled system (\ref{linear1sth})-(\ref{linear1stg}) is given by
  \begin{align}
    &h_{\pm} = \fz^{1+\a}\,\tam{e}^{-\fz/2}\lsb f_\pm\hypF\lb1+\a\pm\b,2+2\a,\fz\rb \right.\nonumber\\
    &\hspace{1cm}\left.+\, u_\pm\,U\lb1+\a\pm\b,2+2\a,\fz\rb\rsb\,,\\
    \nonumber\\
    &g_{\pm} =  \fz^{\a}\,\tam{e}^{-\fz/2}\lsb -f_\pm\,\frac{2\g\,\lb1+2\a\rb}{m} \,\hypF\lb-\a\pm\b,-2\a,\fz\rb\right.\nonumber\\
    &\hspace{1cm}\left. + \, u_\pm\,\frac{\g\,\lb\a\mp\b\rb}{m\,\a}\,U\lb-\a\pm\b,-2\a,\fz\rb\rsb\,,
  \end{align}
  As far as the spectrum is concerned we find two distinct regions: 
  \begin{enumerate}
\item $m^2 > \frac{\l^2 \tilde\m^2}{k}$, i.e. $\fz = \I\abs{\fz}$ (continuum spectrum)\newline
\newline 
In this case none of the above linearly independent solutions are normalizable and analogously to the previous case the boundary conditions at 
$z_0$ are satisfied without further restrictions on the physical mass.  
\item $m^2 < \frac{\l^2 \tilde\m^2}{k}$ (discrete spectrum)\newline
\newline
For real $\fz$ only the $U$-type solution is normalizable since $\hypF\lb\a,\b,\fz\rb\sim\fz^{\a-\b}\,\tam{e}^\fz$ while $U\lb\a,\b,\fz\rb\sim\fz^{-\a}$ as $\fz\to\infty$.
We then set $f_\pm = 0$ and thus a non trivial solution to the boundary conditions (\ref{bc}) requires the vanishing 
of the function
\begin{align}
   &\lb\a+\b\rb\,U\lb1+\a+\b,2+2\a,\fz_0\rb U\lb-\a-\b,-2\a,\fz_0\rb \nonumber\\
   +&\lb\a-\b\rb \,U\lb1+\a-\b,2+2\a,\fz_0\rb U\lb-\a+\b,-2\a,\fz_0\rb \nonumber\,,
   \end{align}
with $\fz_0 = 2 \g z_0\ll 1$. 
\end{enumerate}  
%\newline
Therefore the linear case predicts a finite set of eigenmasses followed by a continuum of eigenvalues. However the mass restriction does not allow 
 a light mode.  
  %%%%%%%%%%%%%%%%%%%%%%%%
%%%%%%%%%%%%%%%%%%%%%%%%
%%%%%%%%%%%%%%%%%%%%%%%%
 \begin{itemize}
 \item Quadratic profile ($a=2$): $\H(z) = \frac{\bar\m}{\sqrt k}\, \frac{z^2}{k^2}$ with $\bar\m = \lb\frac{\sqrt k}{z_0}\rb^2\sqrt{\frac{k}{ z_0 \l_0} + v_0^2 k^3}$.
 \end{itemize}
 %%%%%%%%%%%%%%%%%%%%%%%%
%%%%%%%%%%%%%%%%%%%%%%%%
%%%%%%%%%%%%%%%%%%%%%%%%
In this last case, the (first order) equations of motion turn into
\begin{align}
   & \frac{\di}{\di\,\z}\,h_\pm +  \frac{1}{\z}\lb\a \pm \z\rb h_\pm = - \d\frac{1}{\sqrt\z} \,g_\pm\,,\label{quhpm}\\
   & \frac{\di}{\di\,\z}\,g_\pm -  \frac{1}{\z}\lb\a \pm \z\rb g_\pm =  \d\frac{1}{\sqrt\z}\,h_\pm\,,\label{qugpm}
\end{align}
with $\z = \lb\frac{z}{z_0}\rb^2 \z_0$, $\a = M k$, $\d=\frac{m z_0}{\sqrt{\z_0}}$ and $\z_0 = \frac{\l}{\sqrt k} \sqrt{\frac{k}{  z_0 \l_0} + v_0^2 k^3}$. Redefining the functions as  
\begin{align}
 & h_+ = \z^{-\frac{\a}{2}}\,\tam{e}^{-\frac{1}{2}\z}\,\mathfrak h_+\,,\qquad g_- = \z^{\frac{\a}{2}}\,\tam{e}^{-\frac{1}{2}\z}\,\mathfrak g_-\,,\nonumber\\
 & h_- = \z^{\frac{1+\a}{2}}\,\tam{e}^{-\frac{1}{2}\z}\,\mathfrak h_-\,,\qquad g_+ = \z^{\frac{1-\a}{2}}\,\tam{e}^{-\frac{1}{2}\z}\,\mathfrak g_+\,,\nonumber
\end{align}
the second order differential equations become the following Kummer's equations 
\begin{align}
 &\z\,\mathfrak h''_+ + \lb\frac{1}{2}-\a -\z\rb\mathfrak h'_+ +\frac{\d^2}{4}\mathfrak h_+ =0\,,\\
 &\z\,\mathfrak g''_- + \lb\frac{1}{2}+\a -\z\rb\mathfrak g'_- +\frac{\d^2}{4}\mathfrak g_- =0\,,\\
 &\z\,\mathfrak h''_- + \lb\frac{3}{2}+\a -\z\rb\mathfrak h'_- -\lb 1-\frac{\d^2}{4}\rb\mathfrak h_- =0\,,\\
 &\z\,\mathfrak g''_+ + \lb\frac{3}{2}-\a -\z\rb\mathfrak g'_+ -\lb 1-\frac{\d^2}{4}\rb\mathfrak g_+ =0\,,
 \end{align}
 such that the (normalizable) solutions are
\begin{align}
 & h_+  = a_+ \,\z^{-\frac{\a}{2}}\,\tam{e}^{-\frac{1}{2}\z}\, U\lb- \frac{\d^2}{4},\frac{1}{2}-\a,\z\rb\,,\\
 & h_-  = a_- \,\z^{\frac{1+\a}{2}}\,\tam{e}^{-\frac{1}{2}\z}\, U\lb1- \frac{\d^2}{4},\frac{3}{2}+\a,\z\rb\,,\\
 & g_+  = b_+ \,\z^{\frac{1-\a}{2}}\,\tam{e}^{-\frac{1}{2}\z}\, U\lb1- \frac{\d^2}{4},\frac{3}{2}-\a,\z\rb\,,\\
 & g_-  = b_- \,\z^{\frac{\a}{2}}\,\tam{e}^{-\frac{1}{2}\z}\, U\lb- \frac{\d^2}{4},\frac{1}{2}+\a,\z\rb\,,
 \end{align}
 while the first order constraints impose 
 \be
     b_+ = - \frac{1}{2}\d\,a_+\,,\qquad a_-= \frac{1}{2}\d\,b_-\,,
 \ee
 then the solutions are given by
 \begin{align}
  & \psi_+ = \lb\ba{c} \z^{-\frac{\a}{2}}\,\tam{e}^{-\frac{1}{2}\z}\,U\lb-\frac{\d^2}{4},\frac{1}{2}-\a,\z\rb\,\xi(x)\\
                                  -\frac{\d}{2}\,\z^{\frac{1-\a}{2}}\,\tam{e}^{-\frac{1}{2}\z}\,U\lb1-\frac{\d^2}{4},\frac{3}{2}-\a,\z\rb\,\bar\chi(x)   \ea\rb\,,\\
                                  \nonumber\\
  & \psi_- = \lb\ba{c} \frac{\d}{2}\,\z^{\frac{1+\a}{2}}\,\tam{e}^{-\frac{1}{2}\z}\,U\lb1-\frac{\d^2}{4},\frac{3}{2}+\a,\z\rb\,\xi(x)\\
                                  \z^{\frac{\a}{2}}\,\tam{e}^{-\frac{1}{2}\z}\,U\lb-\frac{\d^2}{4},\frac{1}{2}+\a,\z\rb\,\bar\chi(x)   \ea\rb\,.
 \end{align}
 Finally, a non trivial solution out of the boundary conditions (\ref{bc}) set the restriction 
 \begin{align}
 &U\lb- \frac{\d^2}{4},\frac{1}{2}-\a,\z_0\rb\,U\lb- \frac{\d^2}{4},\frac{1}{2}+\a,\z_0\rb\nonumber\\
 &- \frac{1}{4}\,\d^2\z_0\,U\lb1- \frac{\d^2}{4},\frac{3}{2}-\a,\z_0\rb\,U\lb1- \frac{\d^2}{4},\frac{3}{2}+\a,\z_0\rb = 0\,.\nonumber
 \end{align} 
 This condition quantizes the possible physical masses and indeed it presents 
 a mass gap. To see this notice that the confluent hypergeometric function $U\lb a,b,z\rb$, as a function of $a$, oscillates for negative values of this parameter,
 then we expect to find oscillatory behavior for $\d \gtrsim 2$. Actually, in Fig.~\ref{quant} we plot the above mass quantization condition for $\z_0\sim 10^{-3}$ and two different 
 values of $\a$.

 Further assuming that $\z_0\ll1$ (at the end of section~\ref{phen} we justify this smallness) and for $\abs{a}>1/2$ we find a light smallest Dirac eigenmass, given by
 \be
   m^2 k^2 \approx 2 \,\frac{1}{\G\lb\abs{\a}-\frac{1}{2}\rb}\,\lb\z_0\rb^{\abs{\a}+\frac{1}{2}}\,.\label{light}
 \ee
  For $\abs{\a} < 1/2$ the lowest eigenvalue is not power law suppressed~\cite{Batell:2008me}. Indeed the authors of Ref.~\cite{Batell:2008me} propose such different behavior as a source of the hierarchy. Nevertheless, as we shall see, the hierarchy can be induced within the power law behavior. As for the spectrum is concerned, we are interested in a discrete set of KK excitations, hence from now on we will consider a quadratic profile for the Higgs VEV and as
  pointed out in the previous section, the lightest modes will be identified as the SM fields.\newline
 To finish this section we will present the solutions for the bosonic sector.
   %%%%%%%%%%%%%%%%%%%%%%%%
%%%%%%%%%%%%%%%%%%%%%%%%
%%%%%%%%%%%%%%%%%%%%%%%%
\subsection{\Large Bosonic sector}
\label{EWH}
%%%%%%%%%%%%%%%%%%%%%%%%
%%%%%%%%%%%%%%%%%%%%%%%%
%%%%%%%%%%%%%%%%%%%%%%%%
   %%%%%%%%%%%%%%%%%%%%%%%%
%%%%%%%%%%%%%%%%%%%%%%%%
%%%%%%%%%%%%%%%%%%%%%%%%
\subsubsection{\Large ElectroWeak equations of motion}
%\label{EWH}
%%%%%%%%%%%%%%%%%%%%%%%%
%%%%%%%%%%%%%%%%%%%%%%%%
%%%%%%%%%%%%%%%%%%%%%%%%
 The variational principle applied on the gauge piece of the action (\ref{action}) (taking into account the Higgs VEV) sets the equations of motion
 \be
 \frac{1}{\sqrt{g}}\pd_N \lcb\sqrt{g}\, g^{MR} g^{NS} \pd_{\lsb R\right.} A_{\left.S\rsb}\rcb - g^{MN} g^{RS}  \pd_{\lsb N\right.} A_{\left.R\rsb}\pd_S \phi - g_5^2 \H^2 g^{MN} A_N =0\,,
 \ee
 along with the boundary conditions 
 \be
    F^{5\m}= \left.g^{5 M} g^{\m N} \pd_{\lsb M\right.} A_{\left.N\rsb}\right|_{z_0} =0\,.
 \ee
Applied to the AdS metric (\ref{metric}) in the gauge 
 $A_5 = 0$ and $\pd_M A^M = 0$ we find the equations of motion
 \begin{align}
    &\lb-\Box + \pd_5^2 + \frac{f'}{f} \pd_5 - \phi'\,\pd_5 - g_5^2 \H^2 f^2\rb A^\m =0\,,\\
    \nonumber\\
    &\left.\pd_5 A^\m\right|_{z_0} = 0\,.\label{gaugebc}
 \end{align}
 Decomposing the gauge field as $A^\m (x,z) = v^\m(x) g(z)$ with $\Box v^\m = - m^2 v^\m$ we find 
 \be
 g'' + \frac{f'}{f} g' - \phi' g' - \lb g_5^2 \H^2 f^2 - m^2\rb g = 0\,.
 %\nonumber\\
 %& g'|_{z_0} =0\,.
 \ee
 Then with the redefinitions 
  \be
  \lcb \ba{l}
  g = f^{-1/2}\, \tam{e}^{\frac{1}{2}\phi}\,\tam{e}^{-\frac{1}{2}\th} \,\th^{-\frac{1}{4}}\,\mathfrak g\\
 \th = \sqrt{1+ \frac{g_5^2\bar\m^2}{\m^4 k^3}}\,\m^2 z^2
 \ea\right.
 \ee
 we find
 \be
 \th \frac{\di^2}{\di\th^2}\mathfrak g - \th \frac{\di}{\di\th}\mathfrak g + \mathfrak m^2 \mathfrak g = 0\,,
 %\nonumber\\
 %& g'|_{z_0} =0\,,
 \ee
 where $\mathfrak m = \frac{m}{2\g\m}$ and $\g =\sqrt{1 + \frac{g_5^2 \bar\m^2}{\m^4 k^3}}$. Finding the general solution to this particular case
 of Kummer differential equation is somehow involved. Here we simply give the normalizable solution, although a detailed derivation can be found
 in appendix~\ref{B}. For the gauge field the $m$-th mode is 
 \be
 A^\m (x,z) = v^\m(x)\, \tam{e}^{\frac{1}{2}\lb 1 -\g\rb\m^2 z^2}\,\mathfrak g\lb\g\m^2 z^2\rb\,,
 \ee
 %\nonumber\\
 with  
 \be
 \mathfrak g\lb\th\rb = 1 -\mathfrak m^2 \th\,{\rm ln}\th\,\hypF\lb 1-\mathfrak m^2,2,\th\rb + \frac{\mathfrak m^2\,\th}{\G\lb 1 -\mathfrak m^2\rb}\,\mathfrak u\lb\th\rb\,,
 \ee
 where 
 $$\mathfrak u\lb\th\rb=\sum^\infty_{k=0} \frac{\G\lb1-\mathfrak m^2+k\rb}{k!\,\G\lb2 + k\rb}\lsb\Psi_0\lb 1+k\rb + \Psi_0\lb 2+k\rb-\Psi_0\lb 1-\mathfrak m^2+k\rb\rsb\th^k\,,$$
 and $\Psi_0\lb x\rb \equiv \frac{\G'(x)}{\G(x)}$ is the digamma function. The boundary conditions (\ref{gaugebc}) are
 \be
 \frac{1}{2}\lb 1 -\g\rb\mathfrak g\lb\th_0\rb + \g\, \mathfrak g'\lb\th_0\rb =0\,.
 \ee
Assuming then that $\th_0\ll 1$ we find for the lowest mode ($\mathfrak m^2\ll 1$) 
 \be
 1-\g - 2\g\mathfrak m^2\lsb 1 - \Psi_0\lb2\rb + {\rm ln}\th_0\rsb + \mathcal O \lb\th_0{\rm ln}\th_0\rb = 0\,,
 \ee
 and hence
 \be
   m_W^2 \approx 2 \m^2\g\lb\g-1\rb\frac{1}{\abs{{\rm ln}\th_0}}\,.\label{gauge0mode}
 \ee
 Notice that $\g$ is always greater or equal than 1, and thus the mass is always real.
 On the other hand, when the coupling to the Higgs VEV is switched off ($g_5\to 0$), which is the case of the photon, the lowest eigenmass is zero. 
 Indeed, for $\g = 1$ and $m = 0$ the equations of motion can be written as
 \be
    f\tam{e}^{-\phi}\,\pd_5\lb f\tam{e}^{-\phi}\, \pd_5 A^\m\rb =0\,,
 \ee 
  and together with the boundary condition $f\tam{e}^{-\phi}\,\pd_5 A^\m|_{z_0} =0$, both impose a constant profile for the massless mode.
  %%%%%%%%%%%%%%%%%%%%%%%%
%%%%%%%%%%%%%%%%%%%%%%%%
%%%%%%%%%%%%%%%%%%%%%%%%
\subsubsection{\Large Higgs equations of motion}
%\label{EWH}
%%%%%%%%%%%%%%%%%%%%%%%%
%%%%%%%%%%%%%%%%%%%%%%%%
%%%%%%%%%%%%%%%%%%%%%%%%
  As far as the Higgs is concerned, here we will analyze the perturbations around the background found in section~\ref{higgsvev}, that is $H(x,z) = \H(z) + \tilde H (x,z)$, 
  which gives the free equations of motion   
  \be
    \Box\tilde H-\pd_5^2\tilde H-\lb3\frac{f'}{f} - \phi'\rb\pd_5\tilde H + f^2 m_h^2\,\tilde H =0\,,
\ee
with $\Box \tilde H = -m^2 \tilde H$, while the boundary variation imposes the vanishing of the expression 
\be
\pd_5\tilde H - 2\l_0 k^2 \lsb\lb\abs{\H + \tilde H}^2- v_0^2\rb\tilde H +\abs{\tilde H}^2\H +\lb\tilde H + \tilde H^*\rb\H\rsb\,,
\ee
at the boundary $z=z_0$.
For solving the equations of motion we redefine the the function as $\tilde H = f^{-3/2}\,\t^{\frac{1}{4}}\mathfrak h$ with $\t = \m^2 z^2$ and then we find the Kummer equation
\be
\t\frac{\di^2}{\di\t^2} \mathfrak h + \lb 1  - \t\rb\frac{\di}{\di\t}\mathfrak h + \frac{m^2}{4 \m^2} \mathfrak h =0\,.
\ee
Again we refer to appendix~\ref{B} to find the details of the solution and here we simply write down the normalizable one, which turns out to be 
\begin{align}
 & \tilde H(x,\t) = H(x)\, \G\lb -\frac{m^2}{4\m^2}\rb \t\,{\rm ln}\t \hypF\lb-\frac{m^2}{4\m^2},1,\t\rb\nonumber\\
  &-H(x)\,\t\sum^\infty_{k=0} \frac{\G\lb-\frac{m^2}{4\m^2}+k\rb}{k!\,\G\lb 1+k\rb}\lsb 2 \Psi_0\lb 1+ k\rb- \Psi_0\lb-\frac{m^2}{4\m^2}+k\rb\rsb\t^k\,,
\end{align}
or shortly: $\tilde H(x,z) \equiv H(x)\,\eta\lb\m^2 z^2\rb$.
Assuming that $\t_0 \ll 1$ the boundary conditions can be expressed as
\be
  1 + {\rm ln}\t_0 - 2 \Psi_0 (1) + \frac{\G'\lb-\frac{m^2}{4\m^2}\rb}{\G\lb-\frac{m^2}{4\m^2}\rb} + \mathcal O\lb \t_0\,{\rm ln}\t_0\rb = 0\,,
\ee
and since $\frac{\G'\lb\e\rb}{\G\lb\e\rb}\sim-\frac{1}{\e}$ for $\e\ll 1$ we find that the lowest Higgs mass is 
\be
   m_H^2 \approx 4 \m^2 \frac{1}{\abs{{\rm ln}\t_0}}\,.\label{higgs0mode}
\ee
%%%%%%%%%%%%%%%%%%%%%%%%
%%%%%%%%%%%%%%%%%%%%%%%%
%%%%%%%%%%%%%%%%%%%%%%%%
\section{\Large Phenomenological predictions on the hierarchy of the SM fermion masses}
\label{phen}
%%%%%%%%%%%%%%%%%%%%%%%%
%%%%%%%%%%%%%%%%%%%%%%%%
%%%%%%%%%%%%%%%%%%%%%%%%
 In this section we will apply the previous results to the fermion masses. In particular we will use them as a possible source for the hierarchy of the masses.

  Whenever we have normalizable solutions, a 4d effective action is well defined. Starting from the action (\ref{action}), by means of partial integration 
 and upon the redefinitions (\ref{red}) we find the action (in appendix~\ref{A} we give a proof for the orthogonality of the normalizable solutions)
 \be
   \sum_{n\geq0}\int\di^4x\,\lb\I\,\bar\psi_{\mathfrak f}^{(n)}\spd\psi_{\mathfrak f}^{(n)}-m_{\mathfrak f}^{(n)}\bar\psi_{\mathfrak f}^{(n)}\psi_{\mathfrak f}^{(n)}\rb\,,
 \ee
 where\footnote{Recall that the boundary conditions (\ref{0bc}) set $\xi_+=\xi_-$ and $\chi_+=\chi_-$ and thus we are left with a single Dirac fermion.} 
 $$\psi_{\mathfrak f}^{(n)} \equiv \lb\ba{c}\psi_{\mathfrak f \, L}^{(n)}\\\bar\psi_{\mathfrak f\,R}^{(n)}\ea\rb\,,$$
 with $\mathfrak f$ labeling the fermion flavour. In addition we will assume $z_0 \sim k$.

  Notice, however, that the values of the 5d couplings are physically meaningless by themselves, instead one should consider the effective 4d couplings
  obtained upon the integration over the fifth dimension, and for this aim we need to normalize the wave functions. The normalization constants are given by the expressions
  \begin{align}
     &\lb N^{(m)}_L\rb^{-2}=\frac{k}{2\,\sqrt{\z_0}} \int^\infty_{\z_0}\di\z\,\tam{e}^{-\z}\lsb\z^{-\frac{1}{2}-\a}\,U^2\lb-\frac{\d_m^2}{4},\frac{1}{2}-\a,\z\rb\right.\nonumber\\
                &\qquad\qquad\qquad\qquad\qquad+\frac{\d_m^2}{4} \left.\z^{\frac{1}{2}+\a}\,U^2\lb1-\frac{\d_m^2}{4},\frac{3}{2}+\a,\z\rb\rsb\,,\\
                \nonumber\\
                 &\lb N^{(r)}_R\rb^{-2}=\frac{k}{2\,\sqrt{\z_0}} \int^\infty_{\z_0}\di\z\,\tam{e}^{-\z}\lsb\z^{-\frac{1}{2}+\a}\,U^2\lb-\frac{\d_r^2}{4},\frac{1}{2}+\a,\z\rb\right.\nonumber\\
                &\qquad\qquad\qquad\qquad\qquad+\frac{\d_r^2}{4} \left.\z^{\frac{1}{2}-\a}\,U^2\lb1-\frac{\d_r^2}{4},\frac{3}{2}-\a,\z\rb\rsb\,,
  \end{align}
  for the fermions, while for the Higgs one finds
  \be
     \lb N^{(n)}_H\rb^{-2}=\frac{\lb\t_0\rb^\frac{3}{2}}{2 \m}\int^\infty_{\t_0}\di \t\,\tam{e}^{-\t}\,\t^{-2}\,\eta_{(n)}^2\lb \t\rb\,.
  \ee
   As for the effective Yukawa coupling is concerned,
  we can read it off from the action (\ref{action}) 
  \be
   \sqrt g\,\tam{e}^{-\phi}\l\,H\lb\bar\Psi_L\Psi_R +\bar\Psi_R\Psi_L \rb = \l\,f\,H\lb\bar\psi_+\psi_+ -\bar\psi_-\psi_-\rb\,,
  \ee
  and therefore we find the coupling
  \be
    Y^{(n,m,r)} = y^{(n,m,r)}_{\mathfrak f\,4d}H^{(n)}(x)\lb\psi_{\mathfrak f\,L}^{(m)} (x) \psi_{\mathfrak f\,R}^{(r)} (x) +{\rm h.c}\rb\,,
  \ee
  with 
  \begin{align}
     y^{(n,m,r)}_{\mathfrak f\,4d}& = \frac{k\l\t_0}{4\z_0} N^{(n)}_H N^{(m)}_L N^{(r)}_R \int^\infty_{\z_0} \di\z\,\tam{e}^{-\z} \lb\frac{\t_0}{\z_0}\z\rb^{-1} \eta_{(n)}\lb\frac{\t_0}{\z_0}\z\rb\nonumber\\
    & \hspace{1cm}\times\lsb\d_r\,\z^{\frac{1}{2}+\a}\, U\lb1-\frac{\d_m^2}{4},\frac{3}{2}+\a,\z\rb U\lb-\frac{\d_r^2}{4},\frac{1}{2}+\a,\z\rb\right.\nonumber\\
   &\hspace{1cm}\left. - \d_m\,\z^{\frac{1}{2}-\a}\, U\lb1-\frac{\d_r^2}{4},\frac{3}{2}-\a,\z\rb U\lb-\frac{\d_m^2}{4},\frac{1}{2}-\a,\z\rb\rsb\,,
 \end{align}
 where, for simplicity, we have omitted the flavor index in the integrals. Thus the effective SM Yukawa coupling would be the one between the lightest modes, which will be referred to as simply
 $y^{\mathfrak f}_{4d}\equiv y^{(0,0,0)}_{\mathfrak f\,4d}$.
  In appendix~\ref{C} we give an estimation of the previous hypergeometric integrals based on the smallness of the parameters $\t_0,\z_0$ and the exponential suppression 
  of the integrands. According to that approximation we can estimate the order of magnitude of the fermionic physical (Dirac) mass as
 \be\lcb\ba{c}
    \lb m_{\mathfrak f} \,k\rb^2 \sim \frac{2}{\G\lb\abs{\a_{\mathfrak f}}-\frac{1}{2}\rb}\lsb \sqrt{\frac{\abs{\a_{\mathfrak f}}-\frac{3}{2}}{\abs{\a_{\mathfrak f}}-\frac{1}{2}}}\,\frac{y_{\mathfrak f\,4d}}{\abs{\ln\t_0}}\rsb^{\abs{\a_{\mathfrak f}}+\frac{1}{2}}\,,\qquad \abs{\a}> \frac{3}{2}\\
    \\
    m_{\mathfrak f} \,k \sim \frac{\sqrt 2}{\G\lb\frac{3}{2}-\a\rb
   \lsb\Psi_0\lb\frac{3}{2}-\a\rb + \ln\lb\frac{\t_0}{\z_0}\rb\rsb} \,y_{\mathfrak f\,4d}\,,\qquad \frac{1}{2}< \abs{\a}<\frac{3}{2}
    \ea\right.\label{hierarchy}
    %& m_{\n,\ell}^2\,k^2\approx 2 \lb\frac{\l_{\n,\ell}}{\sqrt k}\sqrt{\frac{k}{z_0 \l_0} + v_0^2 k^3} \rb^{\abs{\a}+\frac{1}{2}}\lb\frac{k}{z_0}\rb^2 \,\frac{1}{\G\lb\abs{\a}-\frac{1}{2}\rb}\,.\label{nulight}
 \ee
 Recall that $\a_{\mathfrak f} = M_{\mathfrak f}\,k$, $M_{\mathfrak f}$ being the 5d fermion bulk mass. 
 
 We will be interested, nevertheless, in the case $\abs{\a}>\frac{3}{2}$. Indeed this power law 
 dependence on the Yukawa coupling allows us to reproduce the hierarchy between the SM fermion masses with a tiny variation in the 5d bulk masses. 
 
 To extract some numbers we will take the scale mass $\m$ associated with the dilaton VEV as $\m\sim$TeV, indeed this is needed for the gauge and the Higgs masses, Eqs.~(\ref{gauge0mode}) and~(\ref{higgs0mode}), respectively,
  to be at the ElectroWeak scale, and similarly we will take the 5d VEV of the Higgs as $v_0 \sim \frac{1}{\sqrt k}$TeV and $\abs{\l_0}\lesssim 1$. For the curvature, however, we will consider two extreme situations:
 \begin{enumerate}
  \item Planckian curvature, $k^{-1} \sim 10^{19}$ GeV.

\begin{figure}[tb]
\begin{center}
\includegraphics[width=8cm]{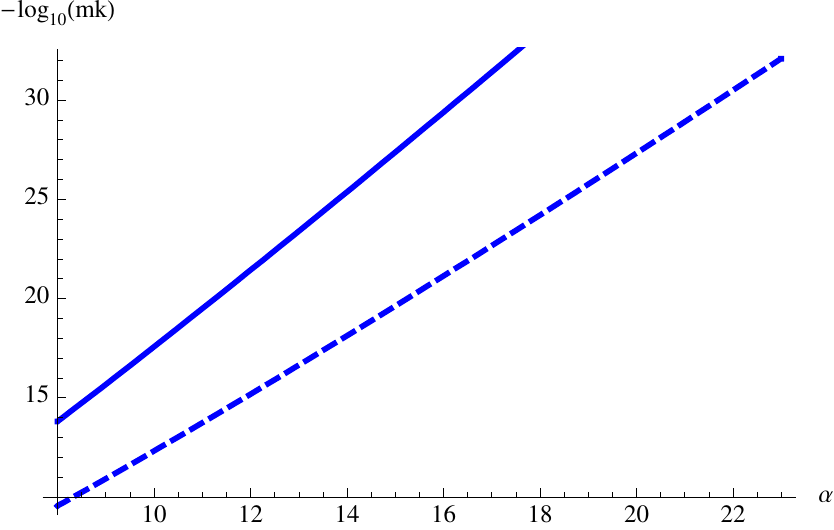}
\caption{\it $-{\rm log}_{10}\lb m k\rb$ as a function of $\a$ for $k^{-1}\sim 10^{19}$GeV and either $y_{4d}\lesssim 1$ (dashed line) or $y_{4d}\sim 0.1$ (solid line).}
\label{planck}
\end{center}
\end{figure}

  As we can see in Fig.~\ref{planck} with order 1 effective Yukawa couplings, $y_{4d} \lesssim 1$, 
 we find the correct order of magnitude for the top quark mass ($m_t k\sim 10^{-17}$) for $ \abs{\a_t} \simeq 13$ and for the neutrinos masses ($m_\n k\sim 10^{-28},10^{-31}$ 
 -depending on wether they are light, $\sim$ eV,  or ultralight $\sim$ meV-) with $ \abs{\a_\n} \simeq 20,\,22$, respectively.  
 
 Of course, the smallness of the mass scale $\m$ compared to the AdS curvature, $\m k\sim 10^{-16}$, and the stabilization of the dilaton VEV deserve an explanation that we are not addressing in this work.
 
 If we, instead, consider slightly suppressed 4d Yukawa couplings, $y_{4d}\sim 0.1$, the suitable values of $\a$ are: $\abs{\a_t} \simeq 10$ and $\abs{\a_\n}\simeq 15,17$. 
 
 \item LHC curvature, $k^{-1} \sim 10^{4}$ GeV.
  
\begin{figure}[tb]
\begin{center}
\includegraphics[width=8cm]{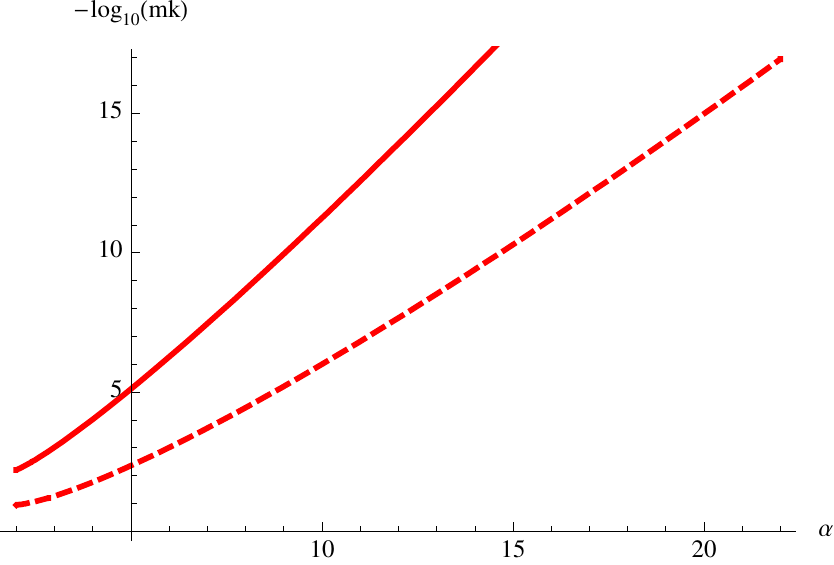}
\caption{\it $-{\rm log}_{10}\lb m k\rb$ as a function of $\a$ for $k^{-1}\sim10^{4}$GeV and either $y_{4d}\lesssim 1$ (dashed line) or $y_{4d}\sim 0.1$ (solid line).}
\label{lhc}
\end{center}
\end{figure}
 In this case (as Fig.~\ref{lhc} shows) with $\mathcal O(1)$ Yukawa couplings we reach the order of magnitude for the top quark ($m_t k \sim 10^{-2}$) with $\abs{\a_t} \simeq 5$ while 
 for the neutrinos ($m_\n k\sim 10^{-13}, 10^{-16}$) we have to consider $\abs{\a_\n}\simeq 18,21$, respectively.
 
 And as before, taking $y_{4d}\sim 0.1$ we instead find $\abs{\a_t} \simeq 2$ and $\abs{\a_\n}\simeq 12,14$.
 \end{enumerate}

 Concerning the parameters $\z^{\mathfrak f}_0,\t_0$, from Eq.~(\ref{light}) we can express $\z^{\mathfrak f}_0$ as  
 \be
 \z^{\mathfrak f}_0 = \lsb \frac{m_{\mathfrak f}^2 k^2}{2}\G\lb\a_{\mathfrak f}-\frac{1}{2}\rb\rsb^{\frac{1}{\a_{\mathfrak f}+\frac{1}{2}}}\,,\nonumber
\ee 
from where we find
\be\nonumber
   \lcb\ba{l} k^{-1} \sim 10^{19}\,{\rm GeV}\,,\t_0\sim10^{-32}\,,\lcb\ba{l} y_{4d} \lesssim 1\,,\qquad \z^{\mathfrak f}_0\sim 10^{-2}\\ y_{4d} \sim 0.1\,,\qquad \z^{\mathfrak f}_0\sim 10^{-3}  \ea\right.\\ 
   \\
                   k^{-1} \sim 10^{4}\,{\rm GeV}\,,\t_0\sim 10^{-2}\,,\lcb\ba{l} y_{4d} \lesssim 1\,,\qquad \z^{\mathfrak f}_0\sim 10^{-1}\\ y_{4d} \sim 0.1\,,\qquad \z^{\mathfrak f}_0\sim 10^{-2}  \ea\right.\ea\right.
\ee

One should compare the previous results with the case of Randall-Sundrum, where the hierarchy of fermions, at least if one forgets about neutrinos, is obtained for $O(1)$ parameters whereas here we need to have $O(10)$ parameters. Although this is not a big difference, the reason why we have such a result is that we are enforcing both the left-handed doublet and right-handed singlet to have the same 5d mass whereas in the case of RS the prescription taken is $c_L=1/2-c_R$. Have we taken a different prescription we would have obtained different results, but we would have lost any possibility of solving the problem analytically. Since the whole point of this paper is just to show how one can generate the hierarchy and not to give a complete model we have preferred to be as simple as possible.

 Let us finished this section by saying that all the masses that we generate in this set-up are \emph{Dirac} since they come from a Yukawa coupling of the Higgs with two different fermions. This will imply that neutrinos are Dirac particles and hence there will be no neutrinoless double beta decay. If it is indeed found that neutrinos are Majorana particles one can think of accommodating that fact into this model by adding some localized new sterile neutrinos in the UV brane or some mass term, also localized, for the right handed partner of the neutrino. We leave this also for further investigation.

      %%%%%%%%%%%%%%%%%%%%%%%%
%%%%%%%%%%%%%%%%%%%%%%%%
%%%%%%%%%%%%%%%%%%%%%%%%
%%%%%%%%%%%%%%%%%%%%%%%%%%%%%%%%%%%%%%%%%%%%%%%%%%%%%%%%%%%%%%%%%%%%%%%%%%%%%%%%%%%%%%
%%%%%%%%%%%%%%%%%%%%%%%%%%%%%%%%%%%%%%%%%%%%%%%%%%%%%%%%%%%%%%%%%%%%%%%%%%%%%%%%%%%%%%
\section{\Large Conclusions}\label{conclusions}
%%%%%%%%%%%%%%%%%%%%%%%%%%%%%%%%%%%%%%%%%%%%%%%%%%%%%%%%%%%%%%%%%%%%%%%%%%%%%%%%%%%%%
%%%%%%%%%%%%%
We have developed a model for ElectroWeak physics embedded in five (non compact) dimensions with an exponentially decaying effective metric such that the fifth dimension
has finite length. The departure of the metric from the AdS case is associated to the VEV acquired by some dilaton field $\phi$ for which we have taken $\langle\phi\rangle = \m^2 z^2$, $z$ representing the fifth coordinate and 
$\m$ being some mass scale,
 although the underlying gravity model giving room to this particular aspect has not been addressed in the current work.
 
In addition, the spacetime has a UV boundary but no IR one. This class of models are commonly known as soft wall models in the literature. A smooth fifth coordinate dependent 
Higgs VEV may be justified within these models, turning out in a power law behavior. Taking this result into account, we have solved the fermionic equations of motion for three different polynomial behaviors 
of the Higgs VEV: constant, linear and quadratic, where the latter is the only  one binding the solutions enough for them to be normalizable and hence yielding a discrete set of KK excitations.

We have further solved the corresponding equations of motion for the Higgs and ElectroWeak sectors and found that the smallest eigenmasses are of the same order of magnitude, both being proportional 
to $\m$ and thus we are forced to consider the latter at the TeV scale.

Finally, we have found that (the order of magnitude of) the lightest fermionic masses behave as a power law of the effective 4d Yukawa couplings, where the exponent is the product between the corresponding 
fermionic 5d bulk mass and the (AdS) radius of curvature ($k$). This power law precisely allows us to reproduce the Standard Model hierarchy between the fermionic masses with a mild 
variation for the bulk masses and even order 1 effective Yukawa couplings. In particular with a bulk 5d mass ranging on $\sim\lb 10,\,20\rb \frac{1}{k}$ we achieve the correct order of magnitude for the top quark
and the neutrinos, respectively, with a Planckian curvature, $k^{-1}\sim 10^{19}$GeV. While for a much lower curvature, $k^{-1}\sim10^4$GeV we reproduce the hierarchy of the masses with 5d bulk masses taking values 
on $\sim\lb2,20\rb\frac{1}{k}$.
We would like to stress again, however, that our main concern here has been to end up with the correct order of magnitude for the SM fermion masses. A more realistic study
incorporating the CKM matrices and the Electroweak constraints, although the main results presented in the current work would 
not substantially change, alike the question concerning the dilaton VEV and its stabilization should be addressed in future investigations.
\newpage
\appendix
%%%%%%%%%%%%%%%%%%%%%%%%
%%%%%%%%%%%%%%%%%%%%%%%%
%%%%%%%%%%%%%%%%%%%%%%%%
\section{Solution to the Kummer equation for integer first parameter ($b$)}\label{B}
%%%%%%%%%%%%%%%%%%%%%%%%
%%%%%%%%%%%%%%%%%%%%%%%%
%%%%%%%%%%%%%%%%%%%%%%%%
%%%%%%%%%%%%%%%%%%%%%%%%
The general form of the Kummer differential equation is~\cite{gradshteyn}
\be
  x\,y'' + \lb b -x\rb y' - a\,y =0\,.
\ee
For $b\notin\mathbb Z$ the two linearly independent solutions are  
\begin{align}
 &y_1= \hypF\lb a,b,x\rb \equiv \frac{\G\lb b\rb}{\G\lb a\rb} \sum^\infty_{k=0} \frac{\G\lb a +k\rb}{k!\,\G\lb b+k\rb} x^k\,,\\
 &y_2 = x^{1-b}\,\hypF\lb 1+a-b,2-b,x\rb\,.
\end{align}
For the confluent hypergeometric function of the first kind, $\hypF\lb a,b,x\rb$, we have 
\be
  \hypF\lb a,b,x\rb\approx x^{a-b}\,\tam{e}^x \lb\frac{\G(b)}{\G(a)} + \mathcal O\lb\frac{1}{x}\rb\rb + x^{-2 a} \lb \frac{(-1)^{a}}{\G\lb b-a\rb} + \mathcal O\lb\frac{1}{x}\rb\rb\,,\nonumber
\ee
thus a non-exponential asymptotic behavior is found with the linear combination\footnote{It behaves as a power law.}
\be
   U\lb a,b,x\rb = \frac{\G\lb 1-b\rb}{\G\lb 1+a-b\rb}\,y_1 + \frac{\G\lb b-1\rb}{\G\lb a\rb}\,y_2\,.
\ee
For $b\in\mathbb Z$ the solutions $y_1,\,y_2$ are no longer independent. Actually one easily checks that 
$$\lim{b}{n}\frac{\G(a)}{\G(b)}\hypF\lb a,b,z\rb\,,$$
does exist for any $n\in \mathbb Z$. Indeed, when $b\to -n$ with $n$ a positive or zero integer one finds
 $$\frac{\G(a)}{\G(b)}\hypF\lb a,b,z\rb\to \frac{\G\lb 1+a + n\rb}{\G\lb 2+n\rb}z^{1+n}\hypF\lb 1+a + n,2+n,z\rb\,,$$
and then the function
\begin{align}
&\tilde U\lb a,b,z\rb \equiv \frac{\G(a)}{\G(b)}\hypF\lb a,b,z\rb - \frac{\G\lb 1+a-b\rb}{\G\lb2-b\rb}z^{1-b}\hypF\lb 1+a-b,2-b,z\rb\nonumber\\
& \qquad\qquad= \frac{\G\lb 1+a-b\rb \G\lb a\rb}{\G\lb 1-b\rb \G\lb b\rb} \,U\lb a,b,z\rb\,,\nonumber
\end{align}
 vanishes as $b$ approaches any integer value. Thus two linearly independent solutions are given by
\begin{align}
& y^n_1 = \lim{b}{n}\frac{\G(a)}{\G(b)}\hypF\lb a,b,z\rb\,,\\
&y^n_2 = \lim{b}{n} \frac{\lb-1\rb^n \pi}{{\rm sin}\pi b}\tilde U\lb a,b,z\rb = \pd_b \left.\tilde U\lb a,b,z\rb\right|_{b=n}\,,
\end{align}
where for the last equality we have used the l'H$\hat{\rm o}$pital rule. In addition notice that $y^n_2$ behaves as a power low when $z\to \infty$ since it is the limit of a function proportional 
to $U\lb a,b,z\rb$.

For the particular cases $b=0$ and $b=1$, corresponding to the gauge and the Higgs cases, respectively, one easily shows that
\begin{align}
&\lim{b}{0}\pd_b \frac{\G(a)}{\G(b)}\hypF\lb a,b,z\rb = \G\lb a\rb - z\sum^\infty_{k=0} \frac{\G\lb 1+a+k\rb}{k!\,\G\lb 2+k\rb}\Psi_0\lb 1+k\rb z^k\,,\nonumber\\
\nonumber\\
&\lim{b}{0}\pd_b\, z^{1-b}\frac{\G\lb1+a-b\rb}{\G\lb2-b\rb} \hypF\lb 1+a-b,2-b,z\rb = \nonumber\\
%\nonumber\\
%&
&z\sum^\infty_{k=0} \frac{\G\lb 1+a+k\rb}{k!\,\G\lb 2+k\rb}\lsb\Psi_0\lb 2+k\rb-\Psi_0\lb 1+a+k\rb\rsb z^k\nonumber\\
&-\G\lb 1+a\rb z\,{\rm ln}z \hypF\lb 1+a,2,z\rb\nonumber\\
\nonumber\\
&\lim{b}{1}\pd_b \frac{\G(a)}{\G(b)}\hypF\lb a,b,z\rb = - \sum^\infty_{k=0} \frac{\G\lb a+k\rb}{k!\,\G\lb 1+k\rb}\Psi_0\lb 1+k\rb z^k\,,\nonumber\\
\nonumber\\
&\lim{b}{1}\pd_b\, z^{1-b}\frac{\G\lb1+a-b\rb}{\G\lb2-b\rb} \hypF\lb 1+a-b,2-b,z\rb = \nonumber\\
%\nonumber\\
%&
&\sum^\infty_{k=0} \frac{\G\lb a+k\rb}{k!\,\G\lb 1+k\rb}\lsb\Psi_0\lb 1+k\rb-\Psi_0\lb a+k\rb\rsb z^k\nonumber\\
&-\G\lb a\rb \,{\rm ln}z \hypF\lb a,1,z\rb\,,\nonumber
\end{align}
with $\Psi_0\lb x\rb \equiv \lcb{\rm ln}\lsb\G\lb x\rb\rsb\rcb'$, the digamma function, and hence we find the (normalizable) solutions presented in subsection~\ref{EWH}. 
%%%%%%%%%%%%%%%%%%%%%%%%
%%%%%%%%%%%%%%%%%%%%%%%%
%%%%%%%%%%%%%%%%%%%%%%%%
\section{Estimate of the hypergeometric integrals}\label{C}
%%%%%%%%%%%%%%%%%%%%%%%%
%%%%%%%%%%%%%%%%%%%%%%%%
%%%%%%%%%%%%%%%%%%%%%%%%
To estimate the value of the wave function normalizations and the effective 4d Yukawa couplings we will assume $\t_0,\z_0\ll 1$. 
Furthermore, since all the integrands are exponentially suppressed we will approximate the hypergeometric functions to the lowest order expansion,
 disregarding the higher values of the integral variable, that is
 \be
   U\lb a,b,x\rb \approx \lsb\frac{\G\lb1-b\rb}{\G\lb1-b+a\rb} + \frac{\G\lb b-1\rb}{\G\lb a\rb}\,x^{1-b}\rsb\lb1 + \mathcal O\lb x\rb\rb\,.
 \ee
 \begin{itemize}
 \item $\abs{\a} >\frac{3}{2}$
 \end{itemize}
 For the case of 
the kinetic normalizations we have
\begin{align}
   &\lb N_R\rb^{-2} = \frac{k}{2\,\sqrt{\z_0}}\frac{\G^2\lb\a-\frac{1}{2}\rb}{\G^2\lb\frac{1}{2}+\a-\frac{\d^2}{4}\rb} \lsb\frac{\d^2}{4} + \frac{\G^2\lb\frac{1}{2}+\a-\frac{\d^2}{4}\rb}{\G^2\lb-\frac{\d^2}{4}\rb}\rsb\nonumber\\
   &\qquad\qquad\times\int^\infty_{\z_0}\di\z\,\tam{e}^{-\z}\,\z^{\frac{1}{2}-\a} \lsb1 + \mathcal O\lb\z^{\a-\frac{1}{2}}\rb\rsb\,.
\end{align}
The leading order integral can be approximated using the L'H$\hat{\rm o}$pital rule when $\z_0\to0$ for which we find
\be
  \int^\infty_{\z_0}\di\z\,\tam{e}^{-\z}\,\z^{\frac{1}{2}-\a} \lsb1 + \mathcal O\lb\z^{\a-\frac{1}{2}}\rb\rsb \approx \frac{\lb\z_0\rb^{\frac{3}{2}-\a}}{\a-\frac{3}{2}}\,\tam{e}^{-\z_0} \lsb 1 + \mathcal O\lb\z_0\rb^{\a -\frac{3}{2}}\rsb\,, 
\ee
and then 
\begin{align}
   %\be
   &\lb N_R\rb^{-2} \approx \frac{k}{2\,\sqrt{\z_0}}\frac{\d^2}{4}\frac{\G^2\lb\a-\frac{1}{2}\rb}{\G^2\lb\frac{1}{2}+\a-\frac{\d^2}{4}\rb}\lsb1 + \frac{\d^2\,\G^2\lb\frac{1}{2}+\a-\frac{\d^2}{4}\rb}{4\,\G^2\lb1-\frac{\d^2}{4}\rb}\rsb \nonumber\\
   &\qquad\qquad\times\frac{\lb\z_0\rb^{\frac{3}{2}-\a}}{\a-\frac{3}{2}}\,\tam{e}^{-\z_0}\,.
   %&\qquad\qquad\times\int^\infty_{\z_0}\di\z\,\tam{e}^{-\z}\,\z^{\frac{1}{2}-\a} \lsb1 + \mathcal O\lb\z^{\a-\frac{1}{2}}\rb\rsb\,.
\end{align}
%\ee
Analogously, for the positive parity spinor $\psi_L$ we find
\begin{align}
   %\be
   &\lb N_L\rb^{-2} \approx \frac{k}{2\,\sqrt{\z_0}}\frac{\G^2\lb\a+\frac{1}{2}\rb}{\G^2\lb\frac{1}{2}+\a-\frac{\d^2}{4}\rb}\lsb1 + \frac{\d^2\,\G^2\lb\frac{1}{2}+\a-\frac{\d^2}{4}\rb}{4\,\G^2\lb1-\frac{\d^2}{4}\rb}\rsb \nonumber\\
   &\qquad\qquad\times\frac{\lb\z_0\rb^{\frac{1}{2}-\a}}{\a-\frac{1}{2}}\,\tam{e}^{-\z_0}\,.
   %&\qquad\qquad\times\int^\infty_{\z_0}\di\z\,\tam{e}^{-\z}\,\z^{\frac{1}{2}-\a} \lsb1 + \mathcal O\lb\z^{\a-\frac{1}{2}}\rb\rsb\,.
\end{align}
Notice that $N_L \lb\a\rb = N_R\lb-\a\rb$ thus if $\a$ were negative the roles of $N_L$ and $N_R$ would simply interchange.

For the Higgs we take $\t^{-1}\eta\lb\t\rb \approx \G\lb-\frac{m_H^2}{4 \m^2}\rb \ln\t$ and hence
\be
  \lb N_H\rb^{-2} \approx \frac{\lb\t_0\rb^{\frac{3}{2}}}{2 \m} \G^2\lb-\frac{m_H^2}{4 \m^2}\rb \G''\lb1\rb\,,
\ee
where we have approximated 
\be
 \int^\infty_{\t_0} \tam{e}^{-\t}\,\ln^2\t \approx  \int^\infty_{0} \tam{e}^{-\t}\,\ln^2\t = \G''\lb1\rb\,. 
\ee
For the effective Yukawa coupling we start from
\begin{align}
  &y_{4d}\approx-\frac{\l k\d \t_0}{4 \z_0} N_L N_R N_H \G\lb-\frac{m_H^2}{4 \m}\rb\frac{\G\lb\a-\frac{1}{2}\rb \G\lb\a+\frac{1}{2}\rb}{\G^2\lb\frac{1}{2}+\a-\frac{\d^2}{4}\rb}\nonumber\\
  &\qquad\times\lsb1 + \frac{\d^2\,\G^2\lb\frac{1}{2}+\a-\frac{\d^2}{4}\rb}{4\,\G^2\lb1-\frac{\d^2}{4}\rb}\rsb\int^\infty_{\z_0} \di\z\,\tam{e}^{-\z}\,\z^{\frac{1}{2}-\a} \ln\lb\frac{\t_0}{\z_0}\z\rb\,,\label{appendixyukawa}
\end{align}
where we have taken 
\begin{align}
  &\z^{\frac{1}{2}+\a}\,U\lb1-\frac{\d^2}{4},\frac{3}{2}+\a,\z\rb U\lb-\frac{\d^2}{4},\frac{1}{2}-\a,\z\rb \nonumber\\
  &- \z^{\frac{1}{2}-\a}\,U\lb1-\frac{\d^2}{4},\frac{3}{2}-\a,\z\rb U\lb-\frac{\d^2}{4},\frac{1}{2}-\a,\z\rb\nonumber\\
   &\approx -\z^{\frac{1}{2}-\a}\,\frac{\G\lb\a-\frac{1}{2}\rb \G\lb\a+\frac{1}{2}\rb}{\G^2\lb\frac{1}{2}+\a-\frac{\d^2}{4}\rb}\lsb1 + \frac{\d^2\,\G^2\lb\frac{1}{2}+\a-\frac{\d^2}{4}\rb}{4\,\G^2\lb1-\frac{\d^2}{4}\rb}\rsb\,.
\end{align}
Using again the L'H$\hat{\rm o}$pital rule we find 
\be
  \int^\infty_{\z_0} \di\z\,\tam{e}^{-\z}\,\z^{\frac{1}{2}-\a} \ln\lb\frac{\t_0}{\z_0}\z\rb\approx\frac{\lb\z_0\rb^{\frac{3}{2}-\a}}{\a-\frac{3}{2}} \tam{e}^{-\z_0} \ln\t_0 \,,
\ee
and by plugging the normalization constants in the expression (\ref{appendixyukawa}) and taking $z_0\sim k$, the former simplifies to 
\be
 y_{4d} \approx \frac{\sqrt 2\,\m k }{\sqrt{\G''\lb1\rb\,\lb\frac{1}{\l_0} + v_0^2 k^3\rb}}\, \z_0\,\abs{\ln\t_0}\,\sqrt{\frac{\a-\frac{1}{2}}{\a-\frac{3}{2}}}\,,\label{appy4d}
\ee
where we have used that $\z_0 = \frac{\l}{\sqrt k} \sqrt{\frac{1}{\l_0} + v_0^2 k^3}$.
In addition, from Eq.~(\ref{light}) we can express $\z_0$ as 
\be
 \z_0 = \lsb \frac{m^2 k^2}{2}\G\lb\a-\frac{1}{2}\rb\rsb^{\frac{1}{\a+\frac{1}{2}}}\,.
\ee 
Finally, since we have set $\m \sim v_0\sqrt k\sim$TeV we may naturally assume
\be
  \frac{\sqrt 2\,\m k}{\sqrt{\G''\lb1\rb\,\lb\frac{1}{\l_0} + v_0^2 k^3\rb}} \sim \mathcal O(1)\,,
\ee
and then, from (\ref{appy4d}) one obtains the corresponding expression in (\ref{hierarchy}). \newline
\begin{itemize}
 \item $\frac{1}{2}<\abs{\a} <\frac{3}{2}$
 \end{itemize}
If $\a$ were such that $1/2<\abs{\a} < 3/2$ we would instead find
\begin{align}
   & \int_{\z_0}^\infty \tam{e}^{-\z} \z^{\frac{1}{2}-\a} \approx \G\lb\frac{3}{2}-\a\rb\,,\nonumber\\
   & \int_{\z_0}^\infty \tam{e}^{-\z} \z^{-\frac{1}{2}-\a} \approx \frac{\lb\z_0\rb^{\frac{1}{2}-\a}}{\a-\frac{1}{2}}\,\tam{e}^{-\z_0} \,,\nonumber\\
   & \int_{\z_0}^\infty \tam{e}^{-\z} \z^{\frac{1}{2}-\a} \ln\lb\frac{\t_0}{\z_0}\z\rb\approx \G\lb\frac{3}{2}-\a\rb
   \lsb\Psi_0\lb\frac{3}{2}-\a\rb + \ln\lb\frac{\t_0}{\z_0}\rb\rsb\,,\nonumber
\end{align}
 with $\Psi_0\lb x\rb$ the digamma function. And finally the effective 4d Yukawa coupling constant turn out to be proportional to the physical mass according to
\be
y_{4d}\approx \frac{1}{\sqrt 2}\G\lb\frac{3}{2}-\a\rb
   \lsb\Psi_0\lb\frac{3}{2}-\a\rb + \ln\lb\frac{\t_0}{\z_0}\rb\rsb\, m k\,.\nonumber
\ee
%%%%%%%%%%%%%%%%%%%%%%%%
%%%%%%%%%%%%%%%%%%%%%%%%
%%%%%%%%%%%%%%%%%%%%%%%%
\section{Some technical aspects concerning the solution to the fermionic equations of motion}\label{A}
%%%%%%%%%%%%%%%%%%%%%%%%
%%%%%%%%%%%%%%%%%%%%%%%%
%%%%%%%%%%%%%%%%%%%%%%%%
The system we are concerned with is of the form
 \begin{align}
    & D_1\,h = g\,,\label{D1}\\
    & D_2\,g = h\,,\label{D2}
 \end{align}
 where $D_{1,2}$ are some (linear) differential operators. The second order (decoupled) system is 
  \begin{align}
    & D_2 D_1\,h = h\,,\label{D21}\\
    & D_1 D_2\,g = g\,.\label{D12}
 \end{align}
 Notice that whenever (\ref{D1}) and (\ref{D21}) hold then (\ref{D2}) and (\ref{D12}) are automatically satisfied and vice versa. 
 In addition, let $\mathcal H = \lcb\left. h:\mathbb R\to\mathbb R\right| D_2 D_1\,h = h\rcb$ and $\G = \lcb\left. g:\mathbb R\to\mathbb R\right| D_1 D_2\,g = g\rcb$ 
be the set of solutions to the second order differential equations (\ref{D21}) and (\ref{D12}), respectively. Then $\forall\,h\in\mathcal H$ ($\forall\,g\in\G$), 
$D_1\,h \in \G$ ($D_2\,g \in \mathcal H$). These considerations are useful when solving the first order constraints.

Concerning the orthonormality of the solutions to the system (\ref{firsth})-(\ref{firstg}) we will show that the operators (\ref{secondh})-(\ref{secondg})
are indeed hermitian with respect to the usual scalar product\footnote{Whenever this scalar product be well defined, which is not the case for the non normalizable solutions.}, say 
$$\langle\phi,\varphi\rangle = \int^\infty_{z_0} \phi^\dagger\varphi\,.$$
 To see this notice that the second order differential operators 
can be compactly written as
\be
\mathcal O H \equiv H'' -\lsb f^2 \lb M + m_D\sig_3\rb^2 - f' \lb M+ m_D\sig_3\rb - f\,m'_D\sig_3\rsb H\,,
\ee  
with $H = \lb h_+,h_-\rb^T$, and analogous expression for $G = \lb g_+,g_-\rb^T$. Then by means of partial integration we find 
\be
\langle\tilde H,\mathcal O H\rangle = \langle\mathcal O \tilde H,H\rangle -\,\left.\tilde H^\dagger H' + \tilde H'^\dagger H\right|_{z_0}\,,
\ee  
and using the first order constraints
\be
H' = -\lb M +m_D\sig_3\rb f\,H - m \,G\,,\qquad \tilde H' = -\lb M +m_D\sig_3\rb f\,\tilde H - m \,\tilde G\,, \nonumber
\ee
the boundary piece reduces to 
\be
        m \, \tilde H^\dagger G - m\, \tilde G^\dagger H\,.
\ee
Finally, the boundary conditions (\ref{bc}) allow us to deduce that $\tilde H^\dagger G = \tilde H^\dagger \sig_1 (-\sig_1) G = - \tilde H^\dagger G$ and thus 
$\langle\tilde H,\mathcal O H\rangle = \langle\mathcal O \tilde H,H\rangle$ for all $\tilde H, H$ solving the system (\ref{firsth})-(\ref{firstg}). The same argument 
leads the the same conclusion for the operators (\ref{secondg}).
%%%%%%%%%%%%%%%%%%%%%%%%
%%%%%%%%%%%%%%%%%%%%%%%%
%%%%%%%%%%%%%%%%%%%%%%%%
\section{Fermionic zero-modes profile}\label{D}
%%%%%%%%%%%%%%%%%%%%%%%%
%%%%%%%%%%%%%%%%%%%%%%%%
%%%%%%%%%%%%%%%%%%%%%%%%
%%%%%%%%%%%%%%%%%%%%%%%%
Here we will plot the profiles of the wave functions corresponding to the top quark, for which $\a\approx 13$ and $\d\sim 10^{-16}$ 

\vspace{1cm}

\begin{figure}[h!]
\begin{center}
\includegraphics[width=7cm]{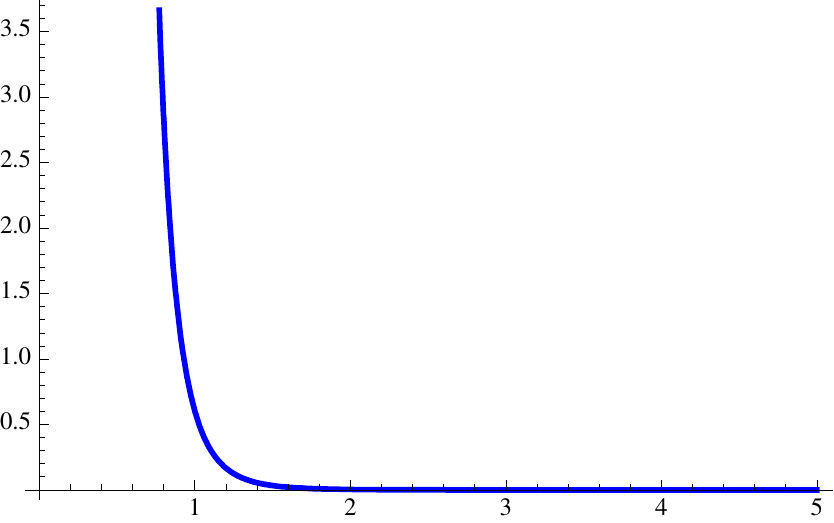}
\hspace{.7cm}
\includegraphics[width=7cm]{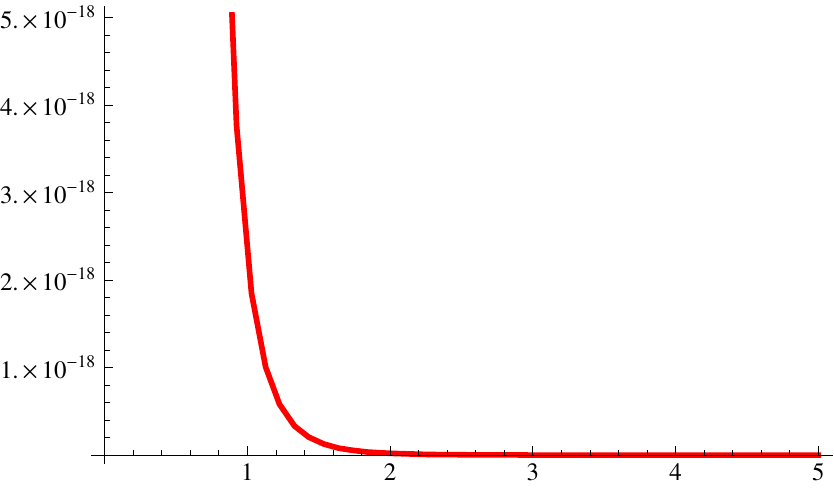}
\\
\vspace{1cm}
\includegraphics[width=7cm]{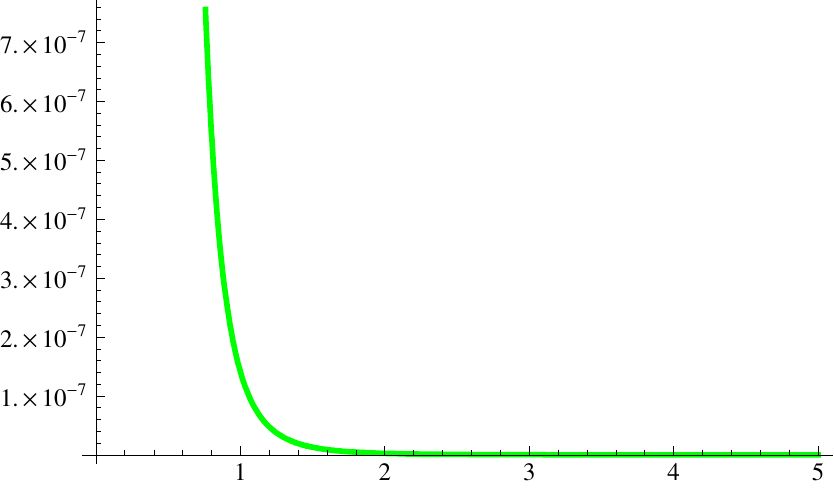}
\hspace{.7cm}
\includegraphics[width=7cm]{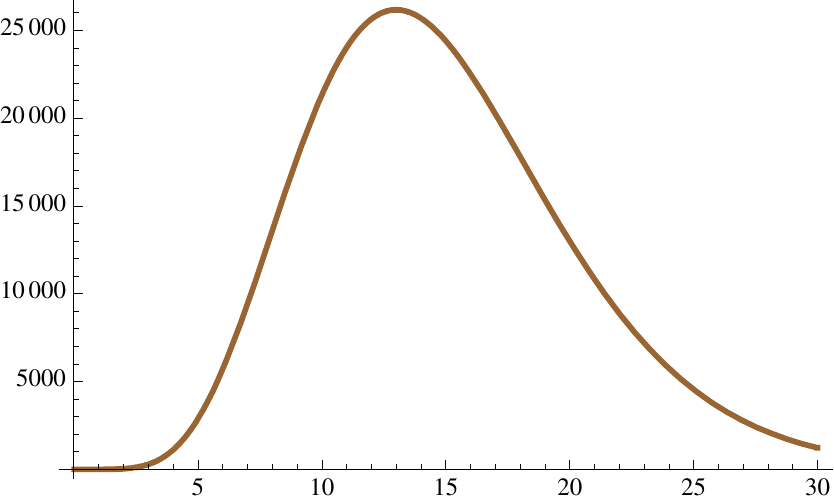}
\caption{\it Plots for the profiles (from left to right and from top to bottom) $h_+$ (blue line), $g_+$ (red line), $h_-$ (green line) and 
$g_-$ (brown line) as a function of $\z$ for $\a = 13$ and $m\,k\sim 10^{-17}$, corresponding to the top quark in the Planckian case.}
%\label{planck}
\end{center}
\end{figure}
As can be seen the profiles are peaked towards the UV brane as opposed to the case of the hard wall scenario (RS) were the top is peaked in the IR brane. For the case of lighter fermions the behavior is qualitatively similar in both scenarios.
%%%%%%%%%%%%%%%%%%%%%%%%
%%%%%%%%%%%%%%%%%%%%%%%%
%%%%%%%%%%%%%%%%%%%%%%%%
%%%%%%%%%%%%%%%%%%%%%%%%
\newpage
%%%%%%%%%%%%%%%%%%%%%%%%
%%%%%%%%%%%%%%%%%%%%%%%%
%%%%%%%%%%%%%%%%%%%%%%%%
%%%%%%%%%%%%%%%%%%%%%%%%
%%%%%%%%%%%%%%%%%%%%%%%%

\end{document}